\pdfoutput=1 

\documentclass[runningheads,envcountsame]{llncs}

\newcommand*{\MainTitle}{Identifiers in Registers}
\newcommand*{\Subtitle}{Describing Network Algorithms with Logic}
\newcommand*{\Title}{\MainTitle{} -- \Subtitle{}}
\newcommand*{\AuthorOne}{Benedikt Bollig}
\newcommand*{\ShortAuthorOne}{B. Bollig}
\newcommand*{\EmailOne}{bollig@lsv.fr}
\newcommand*{\AuthorTwo}{Patricia Bouyer}
\newcommand*{\ShortAuthorTwo}{P. Bouyer}
\newcommand*{\EmailTwo}{bouyer@lsv.fr}
\newcommand*{\AuthorThree}{Fabian Reiter}
\newcommand*{\ShortAuthorThree}{F. Reiter}
\newcommand*{\EmailThree}{fabian.reiter@gmail.com}
\newcommand*{\Authors}{\AuthorOne{}, \AuthorTwo{}, and \AuthorThree{}}
\newcommand*{\Institution}{%
  LSV, CNRS, ENS Paris-Saclay, Universit\'e Paris-Saclay, France}
\newcommand*{\Keywords}{%
  Distributed computing,
  Descriptive complexity,
  Register automata}
\newcommand*{\KnowledgeFootnote}{%
  Built with
  the \href{https://ctan.org/pkg/knowledge}{\texttt{knowledge} package}:
  technical terms are hyperlinked to their definitions.}

\title{\MainTitle{}\thanks{\KnowledgeFootnote}}
\subtitle{\Subtitle}
\titlerunning{\Title}

\author{\AuthorOne{} \and \AuthorTwo{} \and \AuthorThree{}}
\institute{%
  \Institution{} \\
  \email{\EmailOne},\, \email{\EmailTwo},\, \email{\EmailThree}}
\authorrunning{\ShortAuthorOne{}, \ShortAuthorTwo{}, and \ShortAuthorThree{}}

\usepackage{sty/main}        
\usepackage{sty/colors}      
\usepackage{sty/utilities}   
\usepackage{sty/notation}    
\usepackage{sty/terminology} 

\begin{document}

\maketitle

\begin{abstract}
  We propose a formal model of distributed computing
  based on \kl[distributed register automata]{register automata}
  that captures a broad class of synchronous network algorithms.
  The local memory of each process is represented by
  a finite-state controller and a fixed number of \kl{registers},
  each of which can store
  the unique \kl{identifier} of some process in the network.
  To underline the naturalness of our model,
  we show that it has the same expressive power
  as a certain extension of \kl{first-order logic}
  on \kl{graphs} whose \kl{nodes} are equipped with a total order.
  Said extension lets us define new functions on the set of \kl{nodes}
  by means of a so-called \kl{partial fixpoint operator}.
  In spirit,
  our result bears close resemblance to
  a classical theorem of descriptive complexity theory
  that characterizes the complexity class $\PSPACE$
  in terms of partial fixpoint logic
  (a proper superclass of the logic we consider here).
\end{abstract}



\section{Introduction}
\label{sec:introduction}

This paper is part of an ongoing research project aiming to develop
a \emph{descriptive complexity} theory for \emph{distributed computing}.

In classical sequential computing,
descriptive complexity is a well-established field
that connects computational complexity classes
to equi-expressive classes of logical formulas.
It began in the 1970s,
when Fagin showed in~\cite{Fagin74} that
the \kl{graph properties} decidable
by nondeterministic Turing machines in polynomial time
are exactly those definable in existential second-order logic.
This provided a logical---and thus machine-independent---characterization
of the complexity class~$\NPTIME$.
Subsequently,
many other popular classes,
such as~$\PTIME$, $\PSPACE$, and $\EXPTIME$
were characterized in a similar manner
(see for instance the textbooks~%
\cite{DBLP:series/txtcs/GradelKLMSVVW07,DBLP:books/daglib/Immerman99,DBLP:books/sp/Libkin04}).

Of particular interest to us is
a result due to
Abiteboul, Vianu~\cite{DBLP:conf/lics/AbiteboulV89},
and Vardi~\cite{DBLP:conf/stoc/Vardi82},
which states that
on structures equipped with a total order relation,
the properties decidable in $\PSPACE$ coincide with
those definable in \emph{partial fixpoint logic}.
The latter is an extension of first-order logic
with an operator that allows us to inductively define
new relations of arbitrary arity.
Basically,
this means that
new relations can occur as free (second-order) variables
in the logical formulas that define them.
Those variables are initially interpreted as empty relations
and then iteratively updated,
using the defining formulas as update rules.
If the sequence of updates converges to a fixpoint,
then the ultimate interpretations
are the relations reached in the limit.
Otherwise,
the variables are simply interpreted as empty relations.
Hence the term “partial fixpoint”.

While well-developed in the classical case,
descriptive complexity has so far not received much attention
in the setting of distributed network computing.
As far as the authors are aware,
the first step in this direction was taken by Hella~et~al.\
in~\cite{DBLP:conf/podc/HellaJKLLLSV12,DBLP:journals/dc/HellaJKLLLSV15},
where they showed that
basic \emph{modal logic} evaluated on finite graphs
has the same expressive power as
a particular class of \emph{distributed automata}
operating in constant time.
Those automata constitute a weak model of
distributed computing in arbitrary network topologies,
where all \kl{nodes} synchronously execute the same finite-state machine
and communicate with each other
by broadcasting messages to their \kl{neighbors}.
Motivated by this result,
several variants of distributed automata were investigated
by Kuusisto and Reiter
in~\cite{DBLP:conf/csl/Kuusisto13},
\cite{DBLP:conf/icalp/Reiter17}
and~\cite{DBLP:conf/lics/Reiter15}
to establish similar connections with standard logics
such as the \emph{modal $\mu$-calculus}
and \emph{monadic second-order logic}.
However,
since the models of computation investigated in those works
are based on anonymous finite-state machines,
they are much too weak to solve
many of the problems typically considered in distributed computing,
such as leader election or constructing a \kl{spanning tree}.
It would thus be desirable to also characterize stronger models.

A common assumption underlying many distributed algorithms
is that each \kl{node} of the considered network
is given a unique \kl{identifier}.
This allows us, for instance,
to elect a leader
by making the \kl{nodes} broadcast their \kl{identifiers}
and then choose the one with the smallest \kl{identifier} as the leader.
To formalize such algorithms,
we need to go beyond finite-state machines
because the number of bits required to encode a unique \kl{identifier}
grows logarithmically with the number of \kl{nodes} in the network.
Recently,
in~\cite{DBLP:conf/concur/AiswaryaBG15,DBLP:journals/iandc/AiswaryaBG18},
Aiswarya, Bollig and Gastin introduced a synchronous model where,
in addition to a finite-state controller,
\kl{nodes} also have a fixed number of \kl{registers}
in which they can store the \kl{identifiers} of other \kl{nodes}.
Access to those \kl{registers} is rather limited
in the sense that
their contents can be compared
with respect to a total order,
but their numeric values are unknown to the \kl{nodes}.
Similarly,
\kl{register} contents can be copied,
but no new values can be generated.
Since the original motivation for this model was to
automatically verify certain distributed algorithms running on ring networks,
its formal definition is tailored to that particular setting.
However,
the underlying principle can be generalized
to arbitrary networks of unbounded \kl{maximum degree},
which was the starting point for the present work.

\paragraph{Contributions.}

While on an intuitive level,
the idea of finite-state machines equipped with additional \kl{registers}
might seem very natural,
it does not immediately yield
a formal model for distributed algorithms in arbitrary networks.
In particular,
it is not clear
what would be the canonical way for \kl{nodes}
to communicate with a non-constant number of peers,
if we require that
they all follow the same, finitely representable set of rules.

The model we propose here,
dubbed \emph{\kl{distributed register automata}},
is an attempt at a solution.
As in~\cite{DBLP:conf/concur/AiswaryaBG15,DBLP:journals/iandc/AiswaryaBG18},
\kl{nodes} proceed in synchronous rounds
and have a fixed number of \kl{registers},
which they can compare and update without having access to numeric values.
The new key ingredient
that allows us to formalize communication
between \kl{nodes} of unbounded \kl{degree}
is a local computing device we call \emph{\kl{transition maker}}.
This is a special kind of register machine
that the \kl{nodes} can use to scan
the \kl{states} and \kl{register} values of their entire neighborhood
in a sequential manner.
In every round,
each \kl{node} runs the \kl{transition maker}
to update its own \kl{local configuration}
(i.e., its \kl{state} and \kl{register valuation})
based on a snapshot of
the \kl{local configurations} of its \kl{neighbors}
in the previous round.
A way of interpreting this is that the \kl{nodes} communicate
by broadcasting their \kl{local configurations}
as messages to their \kl{neighbors}.
Although the resulting model of computation is by no means universal,
it allows formalizing algorithms for a wide range of problems,
such as
constructing a \kl{spanning tree}
(see Example~\ref{ex:spanning-tree})
or testing whether a \kl{graph} is Hamiltonian
(see Example~\ref{ex:hamiltonian-cycle-automaton}).

Nevertheless,
our model is somewhat arbitrary,
since it could be just one particular choice
among many other similar definitions
capturing different classes of distributed algorithms.
What justifies our choice?
This is where descriptive complexity comes into play.
By identifying a logical formalism
that has the same expressive power as \kl{distributed register automata},
we provide substantial evidence for the naturalness of that model.
Our formalism,
referred to as \emph{\kl{functional fixpoint logic}},
is a fragment of the above-mentioned partial fixpoint logic.
Like the latter,
it also extends \kl{first-order logic}
with a \kl{partial fixpoint operator},
but a weaker one that can only define unary functions
instead of arbitrary relations.
We show that on totally ordered \kl{graphs},
this logic allows one to express precisely
the \kl[graph properties]{properties}
that can be \kl{decided} by \kl{distributed register automata}.
The connection is strongly reminiscent of
Abiteboul, Vianu and Vardi's characterization of $\PSPACE$,
and thus contributes to the broader objective
of extending classical descriptive complexity
to the setting of distributed computing.
Moreover,
given that logical \kl{formulas}
are often more compact and easier to understand
than abstract machines
(compare Examples~\ref{ex:hamiltonian-cycle-automaton}
and~\ref{ex:hamiltonian-cycle-formula}),
logic could also become a useful tool
in the formal specification of distributed algorithms.

The remainder of this paper is structured around our main result:
\begin{theorem}
  \label{thm:main}
  When restricted to finite graphs
  whose nodes are equipped with a total order,
  \kl{distributed register automata}
  are effectively \kl{equivalent} to
  \kl{functional fixpoint logic}.
\end{theorem}
After giving some preliminary definitions in Section~\ref{sec:preliminaries},
we formally introduce
\kl{distributed register automata} in Section~\ref{sec:automata}
and \kl{functional fixpoint logic} in Section~\ref{sec:logic}.
We then sketch the proof of Theorem~\ref{thm:main}
in Section~\ref{sec:automata-vs-logic},
and conclude in Section~\ref{sec:conclusion}.



\section{Preliminaries}
\label{sec:preliminaries}

\AP
We denote
the empty set by~$\intro*\EmptySet$,
the set of nonnegative integers by
$\intro*\Naturals = \set{0,1,2,\dots}$,
and the set of integers by
$\intro*\Integers = \set{\dots,-1,0,1,\dots}$.
The cardinality of any set~$S$ is written as~$\intro*\card{S}$
and the power set as~$\intro*\powerset{S}$.

\AP
In analogy to the commonly used notation for real intervals,
we define the notation
$\intro*\range[m]{n} \defeq \setbuilder{i \in \Integers}{m \leq i \leq n}$
for any $m, n \in \Integers$
such that $m \leq n$.
To indicate that an endpoint is excluded,
we replace the corresponding square bracket
with a parenthesis, e.g.,
$\reintro*\range*[m]{n} \defeq \range[m]{n} \setminus \set{m}$.
Furthermore,
if we omit the first endpoint, it defaults to $0$.
This gives us shorthand notations such as
$\reintro*\range{n} \defeq \range[0]{n}$ and
$\reintro*\range{n}* \defeq \range[0]{n}* = \range[0]{n-1}$.

\AP
All \kl{graphs} we consider are
finite, simple, undirected, and connected.
For notational convenience,
we identify their \kl{nodes} with nonnegative integers,
which also serve as unique \kl{identifiers}.
That is,
when we talk about the \intro{identifier} of a \kl{node},
we mean its numerical representation.
A \intro{graph} is formally represented as a pair
$\Graph = \tuple{\NodeSet, \EdgeSet}$,
where
the set~$\NodeSet$ of \intro{nodes} is equal to~$\range{n}*$,
for some integer $n \geq 2$,
and the set~$\EdgeSet$ consists of undirected \intro{edges}
of the form
$\Edge = \set{\Node[1], \Node[2]} \subseteq \NodeSet$
such that $\Node[1] \neq \Node[2]$.
Additionally,
$\EdgeSet$ must satisfy that
every pair of \kl{nodes}
is connected by a sequence of edges.
The restriction to graphs of size at least two
is for technical reasons;
it ensures that we can always encode Boolean values as \kl{nodes}.

\AP
We refer the reader to~\cite{DBLP:books/daglib/0030488}
for standard graph theoretic terms such as
\intro{neighbor}, \intro{degree}, \intro{maximum degree},
\intro{distance}, and \intro{spanning tree}.

\kl{Graphs} are used to model computer networks,
where \kl{nodes} correspond to processes
and \kl{edges} to communication links.
To represent the current \kl{configuration} of a system
as a \kl{graph},
we equip each \kl{node} with some additional information:
the current \kl{state} of the corresponding process,
taken from a nonempty finite set~$\StateSet$,
and some pointers to other processes,
modeled by a finite set~$\RegisterSet$ of \kl{registers}.

\AP
We call
$\Signature = \tuple{\StateSet, \RegisterSet}$
a \intro{signature}
and define a $\Signature$-\intro{configuration} as a tuple
$\Config = \tuple{\Graph, \StateFunc, \RegisterValFunc}$,
where $\Graph = \tuple{\NodeSet, \EdgeSet}$ is a \kl{graph},
called the \intro{underlying} \kl{graph} of~$\Config$,
$\StateFunc \colon \NodeSet \to \StateSet$
is a \intro{state function}
that assigns to each \kl{node}
a \kl{state}~$\State \in \StateSet$,
and
$\RegisterValFunc \colon \NodeSet \to \NodeSet^{\RegisterSet}$
is a \intro{register valuation function}
that associates with each \kl{node} a \intro{register valuation}
$\RegisterVal \in \NodeSet^{\RegisterSet}$.
The set of all $\Signature$-\kl{configurations}
is denoted by $\intro*\ConfigSet{\Signature}$.
Figure~\ref{fig:local-configuration}
on page~\pageref{fig:local-configuration}
illustrates part of a
$\tuple{
  \set{\State_1, \State_2, \State_3},
  \set{\Register_1, \Register_2, \Register_3}}$-\kl{configuration}.

\AP
If $\RegisterSet = \EmptySet$,
then we are actually dealing with a tuple $\tuple{\Graph,\StateFunc}$,
which we call a \mbox{$\StateSet$-\intro{labeled graph}}.
Accordingly,
the elements of~$\StateSet$
may also be called~\intro{labels}.
A set~$\Property$ of \kl{labeled graphs} will be referred to
as a \intro{graph property}.
Moreover,
if the \kl{labels} are irrelevant,
we set~$\StateSet$ equal to
the singleton~$\intro*\Singleton \defeq \set{\DummyLabel}$,
where~$\intro*\DummyLabel$ is our dummy \kl{label}.
In this case,
we identify $\tuple{\Graph,\StateFunc}$
with~$\Graph$
and call it an \intro{unlabeled} \kl{graph}.



\section{Distributed register automata}
\label{sec:automata}


Many distributed algorithms can be seen as \emph{transducers}.
A leader-election algorithm, for instance,
takes as input a network and outputs the same network,
but with every process storing
the \kl{identifier} of the unique leader
in some dedicated \kl{register}~$\Register$.
Thus,
the algorithm transforms
a $\tuple{\Singleton,\EmptySet}$-\kl{configuration} into
a $\tuple{\Singleton,\{\Register\}}$-\kl{configuration}.
We say that it defines a
\transduction{\tuple{\Singleton,\EmptySet}}{{\tuple{\Singleton,\{\Register\}}}}.
By the same token,
if we consider distributed algorithms
that \emph{\kl{decide}} \kl{graph properties}
(e.g., whether a \kl{graph} is Hamiltonian),
then we are dealing with a
\transduction{\tuple{\InputLabelSet,\EmptySet}}{\tuple{\set{\OutputYes,\OutputNo},\EmptySet}},
where $\InputLabelSet$ is some set of \kl{labels}.
The idea is that a \kl{graph} will be accepted
if and only if
every process eventually outputs $\OutputYes$.

\AP
Let us now formalize the notion of \kl{transduction}.
For any two \kl{signatures}
$\InputSignature = \tuple{\InputLabelSet, \InputRegisterSet}$
and
$\OutputSignature = \tuple{\OutputLabelSet, \OutputRegisterSet}$,
a \intro*\transduction{\InputSignature}{\OutputSignature}
is a \emph{partial} mapping
$\Transduction \colon \ConfigSet{\InputSignature} \to \ConfigSet{\OutputSignature}$
such that, if defined,
$\Transduction\tuple{\Graph, \StateFunc, \RegisterValFunc} =
 \tuple{\Graph, \StateFunc', \RegisterValFunc'}$
for some~$\StateFunc'$ and~$\RegisterValFunc'$.
That is,
a \kl{transduction} does not modify the \kl{underlying} \kl{graph}
but only the \kl{states} and \kl{register valuations}.
We denote the set of all \transduction[s]{\InputSignature}{\OutputSignature}
by $\intro*\TransductionSet{\InputSignature}{\OutputSignature}$
and refer to~$\InputSignature$ and~$\OutputSignature$
as the \intro[input signature]{input} and \intro{output signatures}
of~$\Transduction$.
By extension,
$\InputLabelSet$ and~$\OutputLabelSet$ are called the sets
of \intro[input labels]{input} and \intro{output labels},
and~$\InputRegisterSet$ and~$\OutputRegisterSet$ the sets
of \intro[input registers]{input} and \intro{output registers}.
Similarly,
any $\InputSignature$-\kl{configuration}~$\Config$
can be referred to as an \intro{input configuration} of~$\Transduction$
and~$\Transduction(\Config)$ as an \intro{output configuration}.

\medskip

Next,
we introduce our formal model of distributed algorithms.

\begin{definition}[Distributed register automaton]
  \label{def:automaton}
  \AP
  Let $\InputSignature = \tuple{\InputLabelSet, \InputRegisterSet}$
  and $\OutputSignature = \tuple{\OutputLabelSet, \OutputRegisterSet}$
  be two \kl{signatures}.
  A \intro{distributed register automaton}
  (or simply \reintro{automaton})
  with \kl{input signature}~$\InputSignature$
  and \kl{output signature}~$\OutputSignature$
  is a tuple
  ${\Automaton =
    \tuple{\StateSet, \RegisterSet, \InputFunc, \TransMaker, \HaltingStateSet, \OutputFunc}}$
  consisting of
  a nonempty finite set~$\StateSet$ of \intro{states},
  a finite set~$\RegisterSet$ of \intro{registers}
  that includes both~$\InputRegisterSet$ and~$\OutputRegisterSet$,
  an \intro{input function}
  $\InputFunc \colon \InputLabelSet \to \StateSet$,
  a \kl{transition maker}~$\TransMaker$
  whose specification will be given
  in Definition~\ref{def:transition-maker} below,
  a set~$\HaltingStateSet \subseteq \StateSet$
  of \intro{halting states},
  and an \intro{output function}
  $\OutputFunc \colon \HaltingStateSet \to \OutputLabelSet$.
  The \kl{registers} in
  $\RegisterSet \setminus (\InputRegisterSet \cup \OutputRegisterSet)$
  are called \intro{auxiliary registers}.
\end{definition}

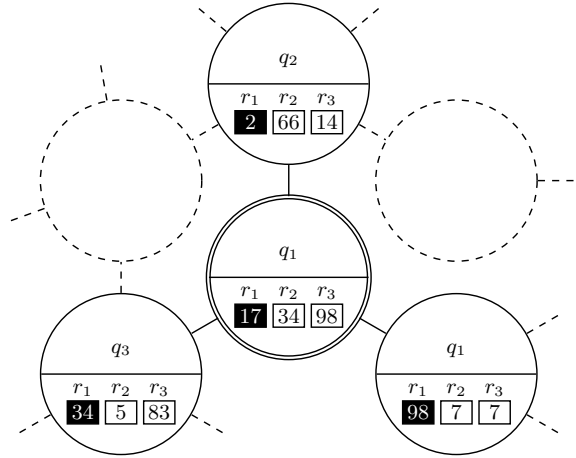
\begin{figure}[tbp]
  \begin{minipage}[c]{0.6\textwidth}
    \scalebox{0.9}{\begin{tikzpicture}[
    semithick,
    vertex/.style = {draw,circle split,inner sep=0ex},
    hidden vertex/.style = {draw,circle,dashed,inner sep=0ex,minimum size=17.5ex},
    state/.style = {minimum height=5ex},
    register valuation/.style = {
      matrix of math nodes, ampersand replacement=\&,
      inner sep=0.5ex, nodes={inner sep=0.5ex}, column sep=0.6ex,
      row 2/.style={nodes={draw,semithick}},
      row 2 column 1/.style={nodes={text=white,fill=black}}
    }
  ]
  \newlength{\outdist}\setlength{\outdist}{12.9ex}
  \tikzbox{\SelfState}{
    \node[state] {$\State_1$};
  }
  \tikzbox{\SelfRegs}{
    \matrix[register valuation] {
      \Register_1 \& \Register_2 \& \Register_3 \\
      17          \& 34          \& 98 \\
    };
  }
  \tikzbox{\NeighborOneState}{
    \node[state] {$\State_2$};
  }
  \tikzbox{\NeighborOneRegs}{
    \matrix[register valuation] {
      \Register_1   \& \Register_2 \& \Register_3 \\
      \sw[c]{2}{99} \& 66          \& 14 \\
    };
  }
  \tikzbox{\NeighborTwoState}{
    \node[state] {$\State_3$};
  }
  \tikzbox{\NeighborTwoRegs}{
    \matrix[register valuation] {
      \Register_1 \& \Register_2   \& \Register_3 \\
      34          \& \sw[c]{5}{99} \& 83 \\
    };
  }
  \tikzbox{\NeighborThreeState}{
    \node[state] {$\State_1$};
  }
  \tikzbox{\NeighborThreeRegs}{
    \matrix[register valuation] {
      \Register_1 \& \Register_2   \& \Register_3 \\
      98          \& \sw[c]{7}{99} \& \sw[c]{7}{99} \\
    };
  }
  \node (Self) [vertex,double,double distance=0.24ex]
        {\usebox{\SelfState} \nodepart{lower} \usebox{\SelfRegs}};
  \node (NeighborOne) at (90:21ex) [vertex]
        {\usebox{\NeighborOneState} \nodepart{lower} \usebox{\NeighborOneRegs}};
  \node (NeighborTwo) at (210:21ex) [vertex]
        {\usebox{\NeighborTwoState} \nodepart{lower} \usebox{\NeighborTwoRegs}};
  \node (NeighborThree) at (330:21ex) [vertex]
        {\usebox{\NeighborThreeState} \nodepart{lower} \usebox{\NeighborThreeRegs}};
  \node (OtherOne) at (30:21ex) [hidden vertex] {};
  \node (OtherTwo) at (150:21ex) [hidden vertex] {};
  \path (NeighborOne)   +( 40:\outdist) coordinate (N1-1)
                        +(140:\outdist) coordinate (N1-2);
  \path (NeighborTwo)   +(210:\outdist) coordinate (N2-1)
                        +(330:\outdist) coordinate (N2-2);
  \path (NeighborThree) +( 30:\outdist) coordinate (N3-1)
                        +(330:\outdist) coordinate (N3-2);
  \path (OtherOne)      +(  0:\outdist) coordinate (O1-1);
  \path (OtherTwo)      +(100:\outdist) coordinate (O2-1)
                        +(200:\outdist) coordinate (O2-2);
  \path[draw]
    (Self) -- (NeighborOne)
    (Self) -- (NeighborTwo)
    (Self) -- (NeighborThree);
  \path[draw,dashed]
    (NeighborOne)   -- (OtherOne)
    (NeighborOne)   -- (OtherTwo)
    (NeighborOne)   -- (N1-1)
    (NeighborOne)   -- (N1-2)
    (NeighborTwo)   -- (OtherTwo)
    (NeighborTwo)   -- (N2-1)
    (NeighborTwo)   -- (N2-2)
    (NeighborThree) -- (N3-1)
    (NeighborThree) -- (N3-2)
    (OtherOne)      -- (O1-1)
    (OtherTwo)      -- (O2-1)
    (OtherTwo)      -- (O2-2);
\end{tikzpicture}}
  \end{minipage}
  \hfill
  \begin{minipage}[c]{0.35\textwidth}
    \caption{
      \protect\RaggedRight
      Part of a \kl{configuration},
      as seen by a single \kl{node}.
      Assuming the \kl{identifiers} of the \kl{nodes}
      are the values represented in black boxes
      (i.e., those stored in \kl{register}~$\Register_1$),
      the \kl{automaton} at \kl{node}~$17$
      will update its own \kl{local configuration}
      $\tuple{
        \State_1,\,
        \set{\Register_1, \Register_2, \Register_3 \mapsto 17,\, 34,\, 98}}$
      by running the \kl{transition maker}
      on the sequence consisting of
      the \kl{local configurations} of \kl{nodes}~$17$, $2$, $34$, and~$98$
      (in that exact order).
    }
    \label{fig:local-configuration}
  \end{minipage}
\end{figure}

\AP
\kl{Automaton}~$\Automaton$ computes a \kl{transduction}
$\TransductionOf{\Automaton}
 \in \TransductionSet{\InputSignature}{\OutputSignature}$.
To do so,
it runs in a sequence of synchronous rounds
on the \kl{input configuration}'s \kl{underlying} \kl{graph}
$\Graph = \tuple{\NodeSet, \EdgeSet}$.
After each round,
the \kl{automaton}'s global configuration
is a $\tuple{\StateSet, \RegisterSet}$-\kl{configuration}
$\Config = \tuple{\Graph, \StateFunc, \RegisterValFunc}$,
i.e., the \kl{underlying} \kl{graph} is always~$\Graph$.
As mentioned before,
for a \kl{node} $\Node \in \NodeSet$,
we interpret $\StateFunc(\Node) \in \StateSet$
as the current \kl{state} of~$\Node$
and $\RegisterValFunc(\Node) \in \NodeSet^\RegisterSet$
as the current \kl{register valuation} of~$\Node$.
Abusing notation,
we let
$\intro*\Local{\Config}{\Node} \defeq
 \tuple{\StateFunc(\Node), \RegisterValFunc(\Node)}$
and say that $\Local{\Config}{\Node}$ is
the \intro{local configuration} of~$\Node$.
In Figure~\ref{fig:local-configuration},
the \kl{local configuration} of \kl{node}~$17$ is
$\tuple{
    \State_1,\,
    \set{\Register_1, \Register_2, \Register_3 \mapsto 17,\, 34,\, 98}}$.

\AP
For a given \kl{input configuration}
$\Config = \tuple{\Graph, \StateFunc, \RegisterValFunc}
 \in \ConfigSet{\InputSignature}$,
the \kl{automaton}'s \intro{initial configuration} is
$\Config' = \tuple{\Graph, \InputFunc \circ \StateFunc, \RegisterValFunc'}$,
where
for all $\Node \in \NodeSet$,
we have
$\RegisterValFunc'(\Node)(\Register) = \RegisterValFunc(\Node)(\Register)$
if $\Register \in \InputRegisterSet$,
and
$\RegisterValFunc'(\Node)(\Register) = \Node$
if $\Register \in \RegisterSet \setminus \InputRegisterSet$.
This means that
every \kl{node}~$\Node$ is initialized to
\kl{state}~$\InputFunc(\StateFunc(\Node))$,
and~$\Node$'s initial \kl{register valuation}~$\RegisterValFunc'(\Node)$
assigns~$\Node$'s own \kl{identifier}
(provided by~$\Graph$)
to all non-\kl{input registers}
while keeping the given values
assigned by~$\RegisterValFunc(\Node)$ to the \kl{input registers}.

\AP
Each subsequent \kl{configuration} is obtained
by running the \kl{transition maker}~$\TransMaker$
synchronously on all \kl{nodes}.
As we will see,
$\TransMaker$ computes a function
\begin{equation*}
  \intro*\Semantics{\TransMaker} \colon
  (\StateSet \times \NodeSet^\RegisterSet)^\KleenePlus \to
  \StateSet \times \NodeSet^\RegisterSet
\end{equation*}
that maps
from nonempty sequences of \kl{local configurations}
to \kl{local configurations}.
This allows the \kl{automaton}~$\Automaton$
to transition
from a given \kl{configuration}~$\Config$
to the next \kl{configuration}~$\Config'$ as follows.
For every \kl{node}~$\Node[1] \in \NodeSet$ of \kl{degree}~$\Degree$,
we consider the list $\Node[2]_1, \dots \Node[2]_\Degree$
of~$\Node[1]$'s \kl{neighbors} sorted in ascending (\kl{identifier}) order,
i.e., $\Node[2]_i < \Node[2]_{i+1}$ for $i \in \range[1]{\Degree}*$.
(See Figure~\ref{fig:local-configuration} for an example,
where~$\Node[1]$ corresponds to \kl{node}~$17$.)
If~$\Node[1]$ is already in a~\kl{halting state},
i.e., if
$\Local{\Config}{\Node[1]} = \tuple{\State, \RegisterVal}
 \in \HaltingStateSet \times \NodeSet^\RegisterSet$,
then its \kl{local configuration} does not change anymore,
i.e., $\Local{\Config'}{\Node[1]} = \Local{\Config}{\Node[1]}$.
Otherwise, we define
$\Local{\Config'}{\Node[1]} =
 \Semantics{\TransMaker}
 \bigl(
   \Local{\Config}{\Node[1]},
   \Local{\Config}{\Node[2]_1}, \dots,
   \Local{\Config}{\Node[2]_\Degree}
 \bigr)$,
which we may write more suggestively as
\begin{equation*}
  \Semantics{\TransMaker} \colon
  \Local{\Config}{\Node[1]}
  \intro*\TransTo{\Local{\Config}{\Node[2]_1}, \dots, \Local{\Config}{\Node[2]_\Degree}}
  \Local{\Config'}{\Node[1]}.
\end{equation*}
Intuitively,
\kl{node}~$\Node[1]$ updates its own \kl{local configuration}
by using~$\TransMaker$ to scan a snapshot
of its \kl{neighbors}' \kl{local configurations}.
As the system is synchronous,
this update procedure is performed simultaneously by all \kl{nodes}.

\AP
A \kl{configuration}
$\Config = \tuple{\Graph, \StateFunc, \RegisterValFunc}$
is called a \intro{halting configuration}
if all \kl{nodes} are in a \kl{halting state},
i.e., if $\StateFunc(\Node) \in \HaltingStateSet$
for all $\Node \in \NodeSet$.
We say that $\Automaton$ \intro{halts}
if it reaches a \kl{halting configuration}.

\AP
The \kl{output configuration}
produced by a \kl{halting configuration}
$\Config = \tuple{\Graph, \StateFunc, \RegisterValFunc}$
is the $\OutputSignature$-\kl{configuration}
$\Config' = \tuple{\Graph, \OutputFunc \circ \StateFunc, \RegisterValFunc'}$,
where
for all $\Node \in \NodeSet$ and $\Register \in \OutputRegisterSet$,
we have
$\RegisterValFunc'(\Node)(\Register) = \RegisterValFunc(\Node)(\Register)$.
In other words,
each \kl{node}~$\Node$
outputs the \kl{state}~$\OutputFunc(\StateFunc(\Node))$
and keeps in its \kl{output registers}
the values assigned by~$\RegisterValFunc(\Node)$.

\AP
It is now obvious that~$\Automaton$ defines a \kl{transduction}
$\intro*\TransductionOf{\Automaton} \colon
 \ConfigSet{\InputSignature} \to \ConfigSet{\OutputSignature}$.
If~$\Automaton$ receives the \kl{input configuration}
$\Config \in \ConfigSet{\InputSignature}$
and eventually \kl{halts} and produces the \kl{output configuration}
$\Config' \in \ConfigSet{\OutputSignature}$,
then $\TransductionOf{\Automaton}(\Config) = \Config'$.
Otherwise (if~$\Automaton$ does not \kl{halt}),
$\TransductionOf{\Automaton}(\Config)$ is undefined.

\paragraph{Deciding \kl{graph properties}.}
\AP
Our primary objective is to use \kl{distributed register automata}
as decision procedures for \kl{graph properties}.
Therefore,
we will focus on \kl{automata}~$\Automaton$
that \kl{halt} in a finite number of rounds
on \emph{every} \kl{input configuration},
and we often restrict to \kl{input signatures} of the form
$\tuple{\InputLabelSet, \EmptySet}$
and the \kl{output signature}
$\tuple{\set{\intro*\OutputYes, \intro*\OutputNo}, \EmptySet}$.
For example,
for $\InputLabelSet = \set{a,b}$,
we may be interested in
the set of $\InputLabelSet$-\kl{labeled graphs}
that have exactly one $a$-labeled \kl{node}~$\Node$ (the “leader”).
We stipulate that~$\Automaton$ \intro{accepts}
an \kl{input configuration}~$\Config$
with \kl{underlying} \kl{graph}
$\Graph = \tuple{\NodeSet, \EdgeSet}$
if $\TransductionOf{\Automaton}(\Config) =
 \tuple{\Graph, \StateFunc, \RegisterValFunc}$
such that $\StateFunc(\Node) = \OutputYes$
for \emph{all} $\Node \in \NodeSet$.
Conversely,
$\Automaton$ \intro{rejects}~$\Config$ if
$\TransductionOf{\Automaton}(\Config) =
 \tuple{\Graph, \StateFunc, \RegisterValFunc}$
such that $\StateFunc(\Node) = \OutputNo$
for \emph{some} $\Node \in \NodeSet$.
This corresponds to the usual definition
chosen in the emerging field of
\emph{distributed decision}~\cite{DBLP:journals/eatcs/FeuilloleyF16}.
Accordingly,
a \kl{graph property}~$\Property$ is \intro{decided} by~$\Automaton$
if the \kl{automaton} \kl{accepts}
all \kl{input configurations} that satisfy~$\Property$
and \kl{rejects} all the others.

\bigskip

It remains to explain
how the~\kl{transition maker}~$\TransMaker$ works internally.

\begin{definition}[Transition maker]
  \label{def:transition-maker}
  \AP
  Suppose that
  $\Automaton =
   \tuple{\StateSet, \RegisterSet, \InputFunc, \TransMaker, \HaltingStateSet, \OutputFunc}$
  is a \kl{distributed register automaton}.
  Then its \intro{transition maker}
  $\TransMaker =
   \tuple{\InnerStateSet, \InnerRegisterSet, \InnerInitState, \InnerTransFunc, \InnerOutputFunc}$
  consists of
  a nonempty finite set~$\InnerStateSet$ of \intro{inner states},
  a finite set~$\InnerRegisterSet$ of \intro{inner registers}
  that is disjoint from~$\RegisterSet$,
  an \intro{inner initial state}~$\InnerInitState \in \InnerStateSet$,
  an \intro{inner transition function}
  $\InnerTransFunc \colon
   \InnerStateSet \times \StateSet \times \powerset{(\InnerRegisterSet \cup \RegisterSet)^2} \!\to
   \InnerStateSet \times (\InnerRegisterSet \cup \RegisterSet)^\InnerRegisterSet$,
  and an \intro{inner output function}
  $\InnerOutputFunc \colon
   \InnerStateSet \to
   \StateSet \times \InnerRegisterSet^\RegisterSet$.
\end{definition}

\AP
Basically,
a \kl{transition maker}
$\TransMaker =
\tuple{\InnerStateSet, \InnerRegisterSet, \InnerInitState, \InnerTransFunc, \InnerOutputFunc}$
is a sequential register automaton
(in the spirit of~\cite{KaminskiF94})
that reads a nonempty sequence
$\tuple{\State_0, \RegisterVal_0}, \dots,
 \tuple{\State_\Degree, \RegisterVal_\Degree}
 \in (\StateSet \times \NodeSet^\RegisterSet)^\KleenePlus$
of \kl{local configurations} of~$\Automaton$
in order to produce a new
\kl{local configuration}~$\tuple{\State', \RegisterVal'}$.
While reading this sequence,
it traverses itself a sequence
$\tuple{\InnerState_0, \InnerRegisterVal_0}, \dots,
 \tuple{\InnerState_{\Degree+1}, \InnerRegisterVal_{\Degree+1}}$
of \intro{inner configurations},
which each consist of
an \kl{inner state}
$\InnerState_i \in \InnerStateSet$
and an \intro{inner register valuation}
$\InnerRegisterVal_i \in (\NodeSet \cup \set{\Undefined})^\InnerRegisterSet$,
where the symbol~$\intro*\Undefined$ represents an undefined value.
For the initial \kl{inner configuration},
we set $\InnerState_0 = \InnerInitState$
and $\InnerRegisterVal_0(\InnerRegister) = \Undefined$
for all $\InnerRegister \in \InnerRegisterSet$.
Now for~$i \in \range{\Degree}$,
when~$\TransMaker$ is in the
\kl{inner configuration}~$\tuple{\InnerState_i, \InnerRegisterVal_i}$
and reads the
\kl{local configuration}~$\tuple{\State_i, \RegisterVal_i}$,
it can compare
all values assigned
to the \kl{inner registers} and \kl{registers}
by~$\InnerRegisterVal_i$ and~$\RegisterVal_i$
(with respect to the order relation on~$\NodeSet$).
In other words,
it has access to the binary relation
${\SmallerRel_i} \subseteq (\InnerRegisterSet \cup \RegisterSet)^2$
such that
for $\InnerRegister[1],\InnerRegister[2] \in \InnerRegisterSet$
and $\Register[1],\Register[2] \in \RegisterSet$,
we have
$\InnerRegister[1] \SmallerRel_i \Register[1]$
if and only if
$\InnerRegisterVal_i(\InnerRegister[1]) < \RegisterVal_i(\Register[1])$,
and analogously for
${\Register[1] \SmallerRel_i \InnerRegister[1]}$,\;
${\InnerRegister[1] \SmallerRel_i \InnerRegister[2]}$,\,
and
${\Register[1] \SmallerRel_i \Register[2]}$.
In particular,
if $\InnerRegisterVal_i(\InnerRegister[1]) = \Undefined$,
then~$\InnerRegister[1]$ is incomparable with respect to~$\SmallerRel_i$.
Equipped with this relation,
$\TransMaker$ transitions
to~$\tuple{\InnerState_{i+1}, \InnerRegisterVal_{i+1}}$
by evaluating
$\InnerTransFunc(\InnerState_i, \State_i, {\SmallerRel_i}) =
 \tuple{\InnerState_{i+1}, \InnerUpdateFunc}$
and computing~$\InnerRegisterVal_{i+1}$ such that
$\InnerRegisterVal_{i+1}(\InnerRegister[1]) = \InnerRegisterVal_i(\InnerRegister[2])$
if $\InnerUpdateFunc(\InnerRegister[1]) = \InnerRegister[2]$,
and
$\InnerRegisterVal_{i+1}(\InnerRegister[1]) = \RegisterVal_i(\Register[2])$
if $\InnerUpdateFunc(\InnerRegister[1]) = \Register[2]$,
where $\InnerRegister[1],\InnerRegister[2] \in \InnerRegisterSet$
and $\Register[2] \in \RegisterSet$.
Finally,
after having read the entire input sequence
and reached the \kl{inner configuration}
$\tuple{\InnerState_{\Degree+1}, \InnerRegisterVal_{\Degree+1}}$,
the \kl{transition maker} outputs the \kl{local configuration}
$\tuple{\State', \RegisterVal'}$
such that
$\InnerOutputFunc(\InnerState_{\Degree+1}) = \tuple{\State', \InnerOutputUpdateFunc}$
and
$\InnerOutputUpdateFunc(\Register) = \InnerRegister$
implies
$\RegisterVal'(\Register) = \InnerRegisterVal_{\Degree+1}(\InnerRegister)$.
Here we assume without loss of generality
that~$\TransMaker$ guarantees that
$\RegisterVal'(\Register) \neq \Undefined$
for all $\Register \in \RegisterSet$.

\begin{remark}
  Recall that
  $\NodeSet = \range{n}*$
  for any \kl{graph}
  $\Graph = \tuple{\NodeSet, \EdgeSet}$
  with~$n$ \kl{nodes}.
  However,
  as \kl{registers} cannot be compared with constants,
  this actually represents an arbitrary assignment
  of unique, totally ordered \kl{identifiers}.
  To determine the smallest \kl{identifier}
  (i.e., $0$),
  the \kl{nodes} can run an algorithm such as the following.
\end{remark}



\begin{example}[Spanning tree]
  \label{ex:spanning-tree}
  We present a simple \kl{automaton}
  $\Automaton =
   \tuple{\StateSet, \RegisterSet, \InputFunc, \TransMaker, \HaltingStateSet, \OutputFunc}$
  with
  \kl{input signature}
  $\InputSignature = \tuple{\Singleton, \EmptySet}$
  and
  \kl{output signature}
  $\OutputSignature = \tuple{\Singleton, \set{\Parent, \Root}}$
  that computes a (breadth-first) \kl{spanning tree}
  of its \kl[input configuration]{input} \kl{graph}
  $\Graph = \tuple{\NodeSet, \EdgeSet}$,
  rooted at the \kl{node} with the smallest \kl{identifier}.
  More precisely,
  in the computed \kl{output configuration}
  $\Config = \tuple{\Graph, \StateFunc, \RegisterValFunc}$,
  every \kl{node} will store
  the \kl{identifier} of its tree parent
  in \kl{register}~$\Parent$
  and the \kl{identifier} of the root
  (i.e., the smallest \kl{identifier})
  in \kl{register}~$\Root$.
  Thus,
  as a side effect,
  $\Automaton$ also solves the leader election problem
  by electing the root as the leader.

  The \kl{automaton} operates in three phases,
  which are represented by the set of \kl{states}
  $\StateSet = \set{1, 2, 3}$.
  A \kl{node} terminates as soon as it reaches the third phase,
  i.e., we set~$\HaltingStateSet = \set{3}$.
  Accordingly,
  the (trivial) \kl[input function]{input} and \kl{output functions} are
  $\InputFunc \colon \DummyLabel \mapsto 1$
  and
  $\OutputFunc \colon 3 \mapsto \DummyLabel$.
  In addition to the \kl{output registers},
  each \kl{node} has an \kl{auxiliary register}~$\Self$
  that will always store its own \kl{identifier}.
  Thus,
  we choose $\RegisterSet = \set{\Self, \Parent, \Root}$.
  For the sake of simplicity,
  we describe the \kl{transition maker}~$\TransMaker$
  in Algorithm~\ref{algo:spanning-tree}
  using pseudocode rules.
  However,
  it should be clear that
  these rules could be relatively easily implemented
  according to Definition~\ref{def:transition-maker}.


\begin{algorithm}[t]
  \caption{\; \kl{Transition maker}
    of the \kl{automaton} from Example~\ref{ex:spanning-tree}}
  \label{algo:spanning-tree}
  $\left.\parbox{33em}{$\begin{array}{@{\quad}l}
      \Keyword{if}~ \,\exists ~\text{\kl{neighbor}}~ \Neighbor \;
      (\Neighbor.\Root < \My.\Root) \colon \\[0.5ex]
      \qquad \My.\CurrentState \gets 1;
      \quad  \My.\Parent \gets \Neighbor.\Self;
      \quad  \My.\Root \gets \Neighbor.\Root
  \end{array}$}\right\}$ \Rule{1} \\[1.5ex]
  $\left.\parbox{33em}{$\begin{array}{@{\quad}l}
      \Keyword{else if}~ \My.\CurrentState = 1 \\[-1ex]
      \phantom{\Keyword{else if}}
        \land\:
        \text{$\forall$ \kl{neighbor} $\Neighbor
          \begin{bmatrix*}[l]
            \, \Neighbor.\Root = \My.\Root \;\land {} \\
            (\Neighbor.\Parent \neq \My.\Self \,\lor\,
              \Neighbor.\CurrentState = 2) \:\!
          \end{bmatrix*}$:
        } \\[-0.5ex]
      \qquad  \My.\CurrentState \gets 2
  \end{array}$}\right\}$ \Rule{2} \\[1.5ex]
  $\left.\parbox{33em}{$\begin{array}{@{\quad}l}
      \Keyword{else if}~
      (\My.\CurrentState = 2 \,\land\, \My.\Root = \My.\Self) \,\lor\,
      (\My.\Parent.\CurrentState = 3) \colon \\[0.5ex]
      \qquad \My.\CurrentState \gets 3
  \end{array}$}\right\}$ \Rule{3} \\[1.5ex]
  $\begin{array}{@{\quad}l}
      \Keyword{else}~ \,\text{do nothing}
  \end{array}$
\end{algorithm}


  All \kl{nodes} start in \kl{state}~$1$,
  which represents the tree-construction phase.
  By \Rule{1},
  whenever an active \kl{node}
  (i.e., a \kl{node} in \kl{state}~$1$ or~$2$)
  sees a \kl{neighbor}
  whose $\Root$ \kl{register} contains a smaller \kl{identifier}
  than the \kl{node}'s own $\Root$ \kl{register},
  it updates its $\Parent$ and $\Root$ \kl{registers} accordingly
  and switches to \kl{state}~$1$.
  To resolve the nondeterminism in \Rule{1},
  we stipulate that~$\Neighbor$ is chosen to be
  the \kl{neighbor} with the smallest \kl{identifier}
  among those whose $\Root$ \kl{register} contains
  the smallest value seen so far.

  As can be easily shown
  by induction on the number of communication rounds,
  the \kl{nodes} have to apply \Rule{1}
  no more than~$\Diameter(\Graph)$ times
  in order for the pointers in \kl{register} $\Parent$
  to represent a valid \kl{spanning tree}
  (where the root points to itself).
  However,
  since the \kl{nodes} do not know
  when~$\Diameter(\Graph)$ rounds have elapsed,
  they must also check that the current \kl{configuration}
  does indeed represent a single tree,
  as opposed to a forest.
  They do so by propagating a signal,
  in form of \kl{state}~$2$,
  from the leaves up to the root.

  By~\Rule{2},
  if an active \kl{node}
  whose \kl{neighbors} all agree on the same root
  realizes that it is a leaf
  or that all of its children are in \kl{state}~$2$,
  then it switches to \kl{state}~$2$ itself.
  Assuming the $\Parent$ pointers in the current \kl{configuration}
  already represent a single tree,
  \Rule{2} ensures that
  the root will eventually be notified of this fact
  (when all of its children are in \kl{state}~$2$).
  Otherwise,
  the $\Parent$ pointers represent a forest,
  and every tree contains at least one \kl{node}
  that has a \kl{neighbor} outside of the tree
  (as we assume the \kl{underlying} \kl{graph} is connected).

  Depending on the \kl[input configuration]{input} \kl{graph},
  a \kl{node} can switch arbitrarily often
  between \kl{states}~$1$ and~$2$.
  Once the \kl{spanning tree} has been constructed
  and every \kl{node} is in \kl{state}~$2$,
  the only \kl{node} that knows this is the root.
  In order for the algorithm to terminate,
  \Rule{3} then makes
  the root broadcast an acknowledgment message down the tree,
  which causes all \kl{nodes} to switch to the \kl{halting state}~$3$.

  (An example run and a proof of correctness
  can be found in Appendix~\ref{app:spanning-tree}.)
  \qed
\end{example}

Building on the \kl{automaton} from Example~\ref{ex:spanning-tree},
we now give an example of a \kl{graph property}
that can be \kl{decided} in our model of distributed computing.
The following \kl{automaton} should be compared to
the logical \kl{formula}
presented later in Example~\ref{ex:hamiltonian-cycle-formula},
which is much more compact and much easier to specify.

\begin{example}[Hamiltonian cycle]
  \label{ex:hamiltonian-cycle-automaton}
  We describe an \kl{automaton}
  with
  \kl{input signature}
  $\InputSignature = \tuple{\Singleton, \set{\Parent, \Root}}$
  and
  \kl{output signature}
  $\OutputSignature = \tuple{\set{\OutputYes, \OutputNo}, \EmptySet}$
  that \kl{decides} if the \kl{underlying} \kl{graph}
  $\Graph = \tuple{\NodeSet, \EdgeSet}$
  of its \kl{input configuration}
  $\Config = \tuple{\Graph, \StateFunc, \RegisterValFunc}$
  is Hamiltonian,
  i.e., whether~$\Graph$ contains a cycle
  that goes through each \kl{node} exactly once.
  The \kl{automaton} works under the assumption
  that~$\RegisterValFunc$ encodes
  a valid \kl{spanning tree} of~$\Graph$
  in the \kl{registers} $\Parent$ and $\Root$,
  as constructed by the \kl{automaton}
  from Example~\ref{ex:spanning-tree}.
  Hence,
  by combining the two \kl{automata},
  we could easily construct a third one
  that \kl{decides} the \kl{graph property} of Hamiltonicity.

  The \kl{automaton}
  $\Automaton =
   \tuple{\StateSet, \RegisterSet, \InputFunc, \TransMaker, \HaltingStateSet, \OutputFunc}$
  presented here
  implements a simple backtracking algorithm
  that tries to traverse~$\Graph$ along a Hamiltonian cycle.
  Its set of \kl{states} is
  $\StateSet =
  \bigl(
    \set{\Unvisited, \Visited, \Backtrack} \times
    \set{\Idle, \Request, \Good, \Bad}
  \bigl)
  \;\cup\;
  \HaltingStateSet$,
  with the set of \kl{halting states}
  $\HaltingStateSet = \set{\OutputYes, \OutputNo}$.
  Each non-\kl{halting state} consists of two components,
  the first one serving for the backtracking procedure
  and the second one for communicating in the \kl{spanning tree}.
  The \kl{input function}~$\InputFunc$ initializes every \kl{node}
  to the \kl{state}~$\tuple{\Unvisited, \Idle}$,
  while the \kl{output function} simply returns
  the answers chosen by the \kl{nodes}, i.e.,
  $\OutputFunc \colon
   \OutputYes \mapsto \OutputYes, \;
   \OutputNo  \mapsto \OutputNo$.
  In addition to the \kl{input registers},
  each \kl{node} has a \kl{register} $\Self$
  storing its own \kl{identifier}
  and a \kl{register} $\Successor$ to point
  to its successor in a (partially constructed) Hamiltonian path.
  That is,
  $\RegisterSet = \set{\Self, \Parent, \Root, \Successor}$.
  We now describe the algorithm in an informal way.
  It is, in principle, easy to implement
  in the \kl{transition maker}~$\TransMaker$,
  but a thorough formalization would be rather cumbersome.

  In the first round,
  the root marks itself as $\Visited$
  and updates its $\Successor$ \kl{register} to point
  towards its smallest \kl{neighbor}
  (the one with the smallest \kl{identifier}).
  Similarly,
  in each subsequent round,
  any $\Unvisited$ \kl{node} that is pointed to
  by one of its \kl{neighbors}
  marks itself as $\Visited$
  and points towards its smallest $\Unvisited$ \kl{neighbor}.
  However,
  if all \kl{neighbors} are already $\Visited$,
  the \kl{node} instead sends the $\Backtrack$ signal to its predecessor
  and switches back to $\Unvisited$
  (in the following round).
  Whenever a $\Visited$ \kl{node} receives the $\Backtrack$ signal
  from its $\Successor$,
  it tries to update its $\Successor$
  to the next-smallest $\Unvisited$ \kl{neighbor}.
  If no such \kl{neighbor} exists,
  it resets its $\Successor$ pointer to itself,
  propagates the $\Backtrack$ signal to its predecessor,
  and becomes $\Unvisited$ in the following round.

  There is only one exception to the above rules:
  if a \kl{node} that is adjacent to the root
  cannot find any $\Unvisited$ \kl{neighbor},
  it chooses the root as its $\Successor$.
  This way,
  the constructed path becomes a cycle.
  In order to check whether that cycle is Hamiltonian,
  the root now broadcast a $\Request$ down the \kl{spanning tree}.
  If the $\Request$ reaches an $\Unvisited$ \kl{node},
  that \kl{node} replies by sending the message $\Bad$
  towards the root.
  On the other hand,
  every $\Visited$ leaf replies with the message $\Good$.
  While $\Bad$ is always forwarded up to the root,
  $\Good$ is only forwarded by \kl{nodes}
  that receive this message from all of their children.
  If the root receives only $\Good$,
  then it knows that the current cycle is Hamiltonian
  and it switches to the \kl{halting state}~$\OutputYes$.
  The information is then broadcast through the entire \kl{graph},
  so that all \kl{nodes} eventually \kl{accept}.
  Otherwise,
  the root sends the $\Backtrack$ signal to its predecessor,
  and the search for a Hamiltonian cycle continues.
  In case there is none
  (in particular, if there is not even an arbitrary cycle),
  the root will eventually receive the $\Backtrack$ signal
  from its greatest \kl{neighbor},
  which indicates that all possibilities have been exhausted.
  If this happens,
  the root switches to the \kl{halting state}~$\OutputNo$,
  and all other \kl{nodes} eventually do the same.
  \qed
\end{example}


\section{Functional fixpoint logic}
\label{sec:logic}

In order to introduce \kl{functional fixpoint logic},
we first give a definition of \kl{first-order logic}
that suits our needs.
\kl{Formulas} will always be evaluated on
\emph{ordered}, undirected, connected, $\InputLabelSet$-\kl{labeled graphs},
where~$\InputLabelSet$ is a fixed finite set of \kl{labels}.

\AP
Throughout this paper,
let~$\intro*\NodeVarSet$ be
an infinite supply of \intro{node variables}
and $\intro*\FuncVarSet$ be
an infinite supply of \intro{function variables};
we refer to them collectively as \intro{variables}.
The corresponding set of \intro{terms}
is generated by the grammar
$\Term \Coloneqq \NodeVar \mid \FuncVar(\Term)$,
where
$\NodeVar \in \NodeVarSet$ and $\FuncVar \in \FuncVarSet$.
%
With this,
the set of \intro{formulas} of \intro{first-order logic}
over~$\InputLabelSet$
is given by the grammar%
\phantomintro{\LABELED}%
\phantomintro{\SMALLER}%
\phantomintro{\LINKED}%
\phantomintro{\NOT}%
\phantomintro{\OR}%
\phantomintro{\EXISTS}%
\begin{equation*}
  \Formula \Coloneqq \reintro*\LABELED{\Label} \Term[2]
                \mid \Term[1] \reintro*\SMALLER \Term[2]
                \mid \Term[1] \reintro*\LINKED  \Term[2]
                \mid \reintro*\NOT \Formula
                \mid \Formula \reintro*\OR \Formula
                \mid \reintro*\EXISTS \NodeVar \, \Formula,
\end{equation*}
where $\Term[1]$ and $\Term[2]$ are \kl{terms},
$\Label \in \InputLabelSet$, and
$\NodeVar \in \NodeVarSet$.
As usual,
we may also use the additional operators
$\intro*\AND$, $\intro*\IMP$, $\intro*\IFF$, $\intro*\FORALL$
to make our \kl{formulas} more readable,
and we define the notations
$\Term[1] \intro*\NOTGREATER \Term[2]$,\,
$\Term[1] \intro*\EQUAL \Term[2]$, and
$\Term[1] \intro*\NOTEQUAL \Term[2]$
as abbreviations for
$\NOT (\Term[2] \SMALLER \Term[1])$,\,
$(\Term[1] \NOTGREATER \Term[2]) \AND
 (\Term[2] \NOTGREATER \Term[1])$, and
$\NOT (\Term[1] \EQUAL \Term[2])$,
respectively.

\AP
The sets of \intro{free variables}
of a \kl{term}~$\Term$ and a \kl{formula}~$\Formula$
are denoted by
$\intro*\free(\Term)$ and $\reintro*\free(\Formula)$,
respectively.
While \kl{node variables} can be bound
by the usual quantifiers~$\EXISTS$ and~$\FORALL$,
\kl{function variables}
can be bound by a \kl{partial fixpoint operator}
that we will introduce below.

\AP
To interpret a \kl{formula}~$\Formula$
on an $\InputLabelSet$-\kl{labeled graph}
$\tuple{\Graph, \StateFunc}$
with $\Graph = \tuple{\NodeSet, \EdgeSet}$,
we are given a \intro{variable assignment}~$\Assignment$
for the \kl{variables} that occur \kl{freely} in~$\Formula$.
This is a partial function
$\Assignment \colon \NodeVarSet \cup \FuncVarSet
 \to \NodeSet \cup \NodeSet^\NodeSet$
such that
$\Assignment(\NodeVar) \in \NodeSet$
if~$\NodeVar$ is a \kl{free} \kl{node variable}
and
$\Assignment(\FuncVar) \in \NodeSet^\NodeSet$
if~$\FuncVar$ is a \kl{free} \kl{function variable}.
We call~$\Assignment(\NodeVar)$ and $\Assignment(\FuncVar)$
the \intro{interpretations}
of~$\NodeVar$ and $\FuncVar$ under~$\Assignment$,
and denote them
by~$\intro*\Interpret{\NodeVar}{\Assignment}$ and
$\reintro*\Interpret{\FuncVar}{\Assignment}$,
respectively.
For a composite \kl{term}~$\Term$,
the corresponding
\kl{interpretation}~$\reintro*\Interpret{\Term}{\Assignment}$
under~$\Assignment$
is defined in the obvious way.

\AP
We write
$\tuple{\Graph, \StateFunc}, \Assignment \intro*\SAT \Formula$
to denote that
$\tuple{\Graph, \StateFunc}$ \intro{satisfies}~$\Formula$
under \kl{assignment}~$\Assignment$.
If~$\Formula$ does not contain any \kl{free variables},
we simply write
$\tuple{\Graph, \StateFunc} \reintro*\SAT \Formula$
and refer to the set~$\Property$ of $\InputLabelSet$-\kl{labeled graphs}
that \kl{satisfy}~$\Formula$
as the \kl{graph property} \intro{defined} by~$\Formula$.
Naturally enough,
we say that
two devices (i.e.,~\kl{automata} or~\kl{formulas})
are \intro{equivalent}
if they specify (i.e.,~\kl{decide} or~\kl{define})
the same \kl{graph property}
and that two classes of devices are equivalent
if their members specify the same class of \kl{graph properties}.

As we assume that the reader is familiar with \kl{first-order logic},
we only define the semantics of the atomic \kl{formulas}
(whose syntax is not completely standard):
\begin{alignat*}{4}
  &\tuple{\Graph, \StateFunc}, \Assignment \,\SAT\, \LABELED{\Label} \Term[2]
  & &\text{iff}
  & &\StateFunc(\Interpret{\Term[2]}{\Assignment}) = \Label
  & & \text{(“$\Term[2]$ has \kl{label} $\Label$”)}, \\
  &\tuple{\Graph, \StateFunc}, \Assignment \,\SAT\, \Term[1] \SMALLER \Term[2]
  & &\text{iff}
  & &\Interpret{\Term[1]}{\Assignment} < \Interpret{\Term[2]}{\Assignment}
  & & \text{(“$\Term[1]$ is smaller than $\Term[2]$”)}, \\
  &\tuple{\Graph, \StateFunc}, \Assignment \,\SAT\, \Term[1] \LINKED \Term[2]
  \qquad & &\text{iff}
  \qquad & &\set{\Interpret{\Term[1]}{\Assignment}, \Interpret{\Term[2]}{\Assignment}}
            \in \EdgeSet
  \qquad & & \text{(“$\Term[1]$ and $\Term[2]$ are adjacent”)}.
\end{alignat*}

\AP
We now turn to \intro{functional fixpoint logic}.
Syntactically,
it is defined as the extension of \kl{first-order logic}
that allows us to write \kl{formulas} of the form
\phantomintro{\PFP}
\phantomintro{\InVar}
\phantomintro{\OutVar}
\begin{equation*}
  \reintro*\PFP \!
  \begin{bmatrix}
    \FuncVar_1 \DEF\:
    \Formula[1]_1(\FuncVar_1, \dots, \FuncVar_{\FuncNum}, \InVar, \OutVar) \\
    \vdots \\
    \FuncVar_{\FuncNum} \DEF\:
    \Formula[1]_{\FuncNum}(\FuncVar_1, \dots, \FuncVar_{\FuncNum}, \InVar, \OutVar)
  \end{bmatrix} \!
  \Formula[2] \,,
  \tag{$\ast$} \label{eq:pfp}
\end{equation*}
where
$\FuncVar_1, \dots, \FuncVar_{\FuncNum} \in \FuncVarSet$,\,
$\reintro*\InVar, \reintro*\OutVar \in \NodeVarSet$,
and $\Formula[1]_1, \dots, \Formula[1]_{\FuncNum}, \Formula[2]$
are \kl{formulas}.
We use the notation
“$\Formula[1]_i(\FuncVar_1, \dots, \FuncVar_{\FuncNum}, \InVar, \OutVar)$”
to emphasize that
$\FuncVar_1, \dots, \FuncVar_{\FuncNum}, \InVar, \OutVar$
may occur \kl{freely} in~$\Formula[1]_i$
(possibly among other \kl{variables}).
The \kl{free variables} of \kl{formula}~\eqref{eq:pfp}~are
given by
$\bigcup_{i \in \range*{\FuncNum}}
 \bigl[
   \free(\Formula[1]_i)
   \setminus \set{\FuncVar_1, \dots, \FuncVar_{\FuncNum}, \InVar, \OutVar}
 \bigr]
 \cup
 \bigl[
   \free(\Formula[2])
   \setminus \set{\FuncVar_1, \dots, \FuncVar_{\FuncNum}}
 \bigr]$.

\AP
The idea is that the \intro{partial fixpoint operator}~$\PFP$
binds the \kl{function variables}
$\FuncVar_1, \dots, \FuncVar_{\FuncNum}$.
The~$\FuncNum$ lines in square brackets
constitute a system of function definitions
that provide an \kl{interpretation} of
$\FuncVar_1, \dots, \FuncVar_{\FuncNum}$,
using the special \kl{node variables}~$\InVar$ and~$\OutVar$
as helpers to represent input and output values.
This is why~$\PFP$ also binds
any \kl{free} occurrences of~$\InVar$ and~$\OutVar$
in $\Formula[1]_1, \dots, \Formula[1]_{\FuncNum}$,
but not~in~$\Formula[2]$.

To specify the semantics of~\eqref{eq:pfp},
we first need to make some preliminary observations.
As before,
we consider a fixed $\InputLabelSet$-\kl{labeled graph}
$\tuple{\Graph, \StateFunc}$
with $\Graph = \tuple{\NodeSet, \EdgeSet}$
and assume that we are given
a \kl{variable assignment}~$\Assignment$
for the \kl{free variables} of~\eqref{eq:pfp}.
With respect to~$\tuple{\Graph, \StateFunc}$ and~$\Assignment$,
each \kl{formula}~$\Formula[1]_i$ induces an operator
$\FuncOperator_{\Formula[1]_i} \colon
 (\NodeSet^\NodeSet)^{\FuncNum} \to \NodeSet^\NodeSet$
that takes some \kl{interpretation} of the \kl{function variables}
$\FuncVar_1, \dots, \FuncVar_{\FuncNum}$
and outputs a new \kl{interpretation} of~$\FuncVar_i$,
corresponding to the function graph defined by~$\Formula[1]_i$
via the \kl{node variables}~$\InVar$ and~$\OutVar$.
For inputs on which~$\Formula[1]_i$ does not define a functional relationship,
the new \kl{interpretation} of~$\FuncVar_i$
behaves like the identity function.
More formally,
given a \kl{variable assignment}~$\ExtendedAssignment$
that extends~$\Assignment$ with \kl{interpretations}
of $\FuncVar_1, \dots, \FuncVar_{\FuncNum}$,
the operator~$\FuncOperator_{\Formula[1]_i}$ maps
$\Interpret{\FuncVar_1}{\ExtendedAssignment}, \dots,
 \Interpret{\FuncVar_{\FuncNum}}{\ExtendedAssignment}$
to the function~$\FuncVarNew_i$
such that
for all $\Node[1] \in \NodeSet$,
\begin{equation*}
  \FuncVarNew_i(\Node[1]) =
  \begin{cases*}
    \Node[2] & if $\Node[2]$ is the unique \kl{node} in $\NodeSet$ s.t.\
               $\tuple{\Graph, \StateFunc},
                \ExtendedAssignment[\InVar,\OutVar \mapsto \Node[1],\Node[2]]
                \,\SAT\, \Formula[1]_i$, \\
    \Node[1] & otherwise.
  \end{cases*}
\end{equation*}
Here,
$\ExtendedAssignment[\InVar,\OutVar \mapsto \Node[1],\Node[2]]$
is the extension of~$\ExtendedAssignment$
\kl{interpreting}~$\InVar$ as~$\Node[1]$ and~$\OutVar$ as~$\Node[2]$.

\AP
In this way,
the operators
$\FuncOperator_{\Formula[1]_1}, \dots, \FuncOperator_{\Formula[1]_{\FuncNum}}$
give rise to an infinite sequence
$\tuple{\FuncVar_1^k, \dots, \FuncVar_{\FuncNum}^k}_{k \geq 0}$
of tuples of functions,
called \intro{stages},
where the initial \kl{stage} contains solely
the identity function~$\intro*\Identity{\NodeSet}$
and
each subsequent \kl{stage} is obtained from its predecessor
by componentwise application of the operators.
More formally,
\begin{equation*}
  \FuncVar_i^0 = \Identity{\NodeSet} =
  \setbuilder{\Node[1] \mapsto \Node[1]}{\Node[1] \in \NodeSet}
  \qquad\text{and}\qquad
  \FuncVar_i^{k + 1} =
  \FuncOperator_{\Formula[1]_i}(\FuncVar_1^k, \dots, \FuncVar_{\FuncNum}^k),
\end{equation*}
for $i \in \range*{\FuncNum}$ and $k \geq 0$.
Now,
since we have not imposed any restrictions
on the \kl{formulas}~$\Formula[1]_i$,
this sequence might never stabilize,
i.e, it is possible that
$\tuple{\FuncVar_1^k, \dots, \FuncVar_{\FuncNum}^k} \neq
 \tuple{\FuncVar_1^{k + 1}, \dots, \FuncVar_{\FuncNum}^{k + 1}}$
for all $k \geq 0$.
Otherwise,
the sequence reaches a (simultaneous) fixpoint
at some position~$k$ no greater than
$\card{\NodeSet}^{\card{\NodeSet} \cdot \FuncNum}$
(the number of $\FuncNum$-tuples of functions on~$\NodeSet$).

\AP
We define the \intro{partial fixpoint}
$\tuple{\FuncVar_1^\infty, \dots, \FuncVar_{\FuncNum}^\infty}$
of the operators
$\FuncOperator_{\Formula[1]_1}, \dots, \FuncOperator_{\Formula[1]_{\FuncNum}}$
to be the reached fixpoint if it exists,
and the tuple of identity functions otherwise.
That is,
for $i \in \range*{\FuncNum}$,
\begin{equation*}
  \FuncVar_i^\infty =
  \begin{cases*}
    \FuncVar_i^k
    & if there exists $k \geq 0$ such that
      $\FuncVar_j^k = \FuncVar_j^{k + 1}$ for all $j \in \range*{\FuncNum}$, \\
    \Identity{\NodeSet}
    & otherwise.
  \end{cases*}
\end{equation*}

Having introduced the necessary background,
we can finally provide the semantics
of the \kl{formula}
$\PFP [\FuncVar_i \DEF \Formula[1]_i]_{i \in \range*{\FuncNum}} \, \Formula[2]$
presented in~\eqref{eq:pfp}:
\begin{equation*}
  \tuple{\Graph, \StateFunc}, \Assignment
  \:\SAT\:
  \PFP [\FuncVar_i \DEF \Formula[1]_i]_{i \in \range*{\FuncNum}} \, \Formula[2]
  \qquad \text{iff} \qquad
  \tuple{\Graph, \StateFunc},
  \Assignment[\FuncVar_i \mapsto \FuncVar_i^\infty]_{i \in \range*{\FuncNum}}
  \:\SAT\:
  \Formula[2],
\end{equation*}
where
$\Assignment[\FuncVar_i \mapsto \FuncVar_i^\infty]_{i \in \range*{\FuncNum}}$
is the extension of~$\Assignment$
that \kl{interprets}~$\FuncVar_i$ as~$\FuncVar_i^\infty$,
for $i \in \range*{\FuncNum}$.
In other words,
the \kl{formula}
$\PFP [\FuncVar_i \DEF \Formula[1]_i]_{i \in \range*{\FuncNum}} \, \Formula[2]$
can intuitively be read as
\begin{equation*}
  \text{
    “if $\FuncVar_1, \dots, \FuncVar_{\FuncNum}$
    are \kl{interpreted} as the \kl{partial fixpoint} of
    $\Formula[1]_1, \dots, \Formula[1]_{\FuncNum}$,
    then~$\Formula[2]$ holds”.
  }
\end{equation*}

\subsection*{Syntactic sugar}
\label{ssec:syntactic-sugar}

Before we consider a concrete \kl{formula}
(in Example~\ref{ex:hamiltonian-cycle-formula}),
we first introduce some “syntactic sugar”
to make using \kl{functional fixpoint logic} more pleasant.

\paragraph{Set variables.}
\phantomintro{\NOTIN}
According to our definition of \kl{functional fixpoint logic},
the operator~$\PFP$ can bind only \kl{function variables}.
However,
functions can be used to encode sets of \kl{nodes}
in a straightforward manner:
any set~$\NodeSubset$ may be represented by a function
that maps
\kl{nodes} outside of~$\NodeSubset$ to themselves
and
\kl{nodes} inside~$\NodeSubset$ to \kl{nodes} distinct from themselves.
Therefore,
we may fix an infinite supply~$\intro*\SetVarSet$
of \intro{set variables},
and extend the syntax of \kl{first-order logic}
to allow atomic \kl{formulas} of the form
$\Term \intro*\IN \SetVar$,
where~$\Term$ is a \kl{term}
and~$\SetVar$ is a \kl{set variable} in~$\SetVarSet$.
Naturally,
the semantics is that
“$\Term$~is an element of~$\SetVar$”.
To bind \kl{set variables},
we can then write
\kl[partial fixpoint operator]{partial fixpoint} \kl{formulas}
of the form
$\PFP
 \bigl[
   \tuple{\FuncVar_i \DEF \Formula[1]_i}_{i \in \range*{\FuncNum}},
   \tuple{\SetVar_i \DEF \Formula[3]_i}_{i \in \range*{m}}
 \bigr] \,
 \Formula[2]$,
where
$\FuncVar_1, \dots, \FuncVar_{\FuncNum} \in \FuncVarSet$,\,
$\SetVar_1, \dots, \SetVar_m \in \SetVarSet$,
and
$\Formula[1]_1, \dots, \Formula[1]_{\FuncNum},
 \Formula[3]_1, \dots, \Formula[3]_m, \Formula[2]$
are \kl{formulas}.
The \kl{stages} of the \kl{partial fixpoint} induction
are computed as before,
but each \kl{set variable}~$\SetVar_i$ is initialized to~$\EmptySet$,
and falls back to~$\EmptySet$
in case the sequence of \kl{stages} does not converge to a fixpoint.
(More details can be found in Appendix~\ref{ssec:sets-as-functions}.)

\paragraph{Quantifiers over functions and sets.}
\phantomintro{\FORALLSETS}
\kl{Partial fixpoint} inductions allow us
to iterate over various \kl{interpretations}
of \kl[function variables]{function} and \kl{set variables}
and thus provide a way of expressing
(second-order) quantification over functions and sets.
Since we restrict ourselves to
\kl{graphs} whose \kl{nodes} are totally ordered,
we can easily define a suitable order of iteration
and a corresponding \kl{partial fixpoint} induction
that traverses all possible \kl{interpretations}
of a given \kl[function variable]{function} or \kl{set variable}.
To make this more convenient,
we enrich the language of \kl{functional fixpoint logic}
with second-order quantifiers,
allowing us to write \kl{formulas} of the form
$\intro*\EXISTSFUNC \FuncVar \, \Formula[1]$
and
$\intro*\EXISTSSET \SetVar \, \Formula[1]$,
where
$\FuncVar \in \FuncVarSet$, $\SetVar \in \SetVarSet$,
and~$\Formula[1]$ is a \kl{formula}.
The semantics is the standard one.
(More details can be found in Appendix~\ref{ssec:function-quantifiers}.)

\smallskip

As a consequence,
it is possible to express
any \kl{graph property} definable in \emph{monadic second-order logic},
the extension of \kl{first-order logic} with set quantifiers.

\begin{corollary}
  When restricted to finite \kl{graphs}
  equipped with a total order,
  \kl{functional fixpoint logic} is strictly more expressive
  than monadic second-order logic.
\end{corollary}

The strictness of the inclusion in the above corollary
follows from the fact that
even on totally ordered \kl{graphs},
Hamiltonicity cannot be \kl{defined} in monadic second-order logic
(see, e.g., the proof in \cite[Prp.~5.13]{DBLP:books/daglib/0030804}).
As the following example shows,
this \kl[graph property]{property} is easy to express
in \kl{functional fixpoint logic}.

\begin{example}[Hamiltonian cycle]
  \label{ex:hamiltonian-cycle-formula}
  The following \kl{formula} of \kl{functional fixpoint logic}
  \kl{defines} the \kl{graph property} of Hamiltonicity.
  That is,
  an \kl{unlabeled} \kl{graph}~$\Graph$
  \kl{satisfies} this \kl{formula}
  if and only if
  there exists a cycle in~$\Graph$
  that goes through each \kl{node} exactly once.
  \begin{equation*}
    \EXISTSFUNC \FuncVar
    \begin{bmatrix*}[l]
      \;\FORALL \NodeVar[1]
      \bigl(
        \FuncVar(\NodeVar[1]) \LINKED \NodeVar[1]
      \bigr)
      \;\AND\;
      \FORALL \NodeVar[1] \, \EXISTS \NodeVar[2]
      \bigl[
        \FuncVar(\NodeVar[2]) \EQUAL \NodeVar[1]
        \,\AND\,
        \FORALL \NodeVar[3]
        \bigl(
          \FuncVar(\NodeVar[3]) \EQUAL \NodeVar[1]
          \,\IMP\,
          \NodeVar[3] \EQUAL \NodeVar[2]
        \bigr)
      \bigr] \;\AND\, {} \\[1ex]
      \;\FORALLSETS \SetVar
      \Bigl(
        \bigl[
          \EXISTS \NodeVar[1] (\NodeVar[1] \IN \SetVar)
          \,\AND\,
          \FORALL \NodeVar[2]
          \bigl(
            \NodeVar[2] \IN \SetVar \IMP \FuncVar(\NodeVar[2]) \IN \SetVar
          \bigr)
        \bigr]
        \:\IMP\:
        \FORALL \NodeVar[2] (\NodeVar[2] \IN \SetVar)
      \Bigr)
    \end{bmatrix*}
  \end{equation*}
  Here,
  $\NodeVar[1], \NodeVar[2], \NodeVar[3] \in \NodeVarSet$,\,
  $\SetVar \in \SetVarSet$, and
  $\FuncVar \in \FuncVarSet$.
  Intuitively,
  we represent a given Hamiltonian cycle
  by a function~$\FuncVar$
  that tells us for each \kl{node}~$\NodeVar[1]$,
  which of $\NodeVar[1]$'s \kl{neighbors} we should visit next
  in order to traverse the entire cycle.
  Thus,
  $\FuncVar$ actually represents a directed version of the cycle.

  To ensure the existence of a Hamiltonian cycle,
  our \kl{formula} states
  that there is a function~$\FuncVar$
  satisfying the following two conditions.
  By the first line,
  each \kl{node}~$\NodeVar[1]$ must have exactly
  one $\FuncVar$-predecessor and
  one $\FuncVar$-successor,
  both of which must be \kl{neighbors} of~$\NodeVar[1]$.
  By the second line,
  if we start at any \kl{node}~$\NodeVar[1]$
  and collect into a set~$\SetVar$
  all the \kl{nodes} reachable from~$\NodeVar[1]$
  (by following the path specified by~$\FuncVar$),
  then~$\SetVar$ must contain all \kl{nodes}.
  \qed
\end{example}




\section{Translating between automata and logic}
\label{sec:automata-vs-logic}

Having introduced both \kl{automata}
and \kl[functional fixpoint logic]{logic},
we can proceed to explain the first part of Theorem~\ref{thm:main}
(stated in Section~\ref{sec:introduction}),
i.e., how \kl{distributed register automata}
can be translated into \kl{functional fixpoint logic}
(see Appendix~\ref{app:automata-to-logic} for a full proof).

\begin{proposition}
  \label{prp:automata-to-logic}
  For every \kl{distributed register automaton}
  that \kl{decides} a \kl{graph property},
  we can construct
  an \kl{equivalent} \kl{formula} of \kl{functional fixpoint logic}.
\end{proposition}

\begin{proof}[sketch]
  Given a \kl{distributed register automaton}
  $\Automaton =
  \tuple{\StateSet, \RegisterSet, \InputFunc,
    \TransMaker, \HaltingStateSet, \OutputFunc}$
  \kl{deciding} a \kl{graph property}~$\Property$
  over \kl{label} set~$\InputLabelSet$,
  we can construct a \kl{formula}~$\Formula[1]_{\Automaton}$
  of \kl{functional fixpoint logic}
  that \kl{defines}~$\Property$.
  For each \kl{state}~$\State \in \StateSet$,
  our \kl{formula} uses a \kl{set variable}~$\SetVar[1]_{\State}$
  to represent the set of \kl{nodes} of the input \kl{graph}
  that are in \kl{state}~$\State$.
  Also,
  for each \kl{register}~$\Register \in \RegisterSet$,
  it uses a \kl{function variable}~$\FuncVar[1]_{\Register}$
  to represent the function that maps each \kl{node}~$\Node[1]$
  to the \kl{node}~$\Node[2]$
  whose \kl{identifier} is stored in~$\Node[1]$'s \kl{register}~$\Register$.
  By means of a \kl{partial fixpoint operator},
  we enforce that
  on any $\InputLabelSet$-\kl{labeled graph} $\tuple{\Graph, \StateFunc}$,
  the final \kl{interpretations}
  of~$\tuple{\SetVar[1]_{\State}}_{\State \in \StateSet}$
  and~$\tuple{\FuncVar[1]_{\Register}}_{\Register \in \RegisterSet}$
  represent the \kl{halting configuration}
  reached by~$\Automaton$ on~$\tuple{\Graph, \StateFunc}$.
  The main \kl{formula} is simply
  \begin{equation*}
    \Formula[1]_{\Automaton} \defeq\:
    \PFP \!
    \begin{bmatrix}
      \tuple{
        \SetVar[1]_{\State[2]} \DEF \Formula[1]_{\State[2]}
      }_{\State[2] \in \StateSet} \\
      \tuple{
        \FuncVar[1]_{\Register} \DEF \Formula[1]_{\Register}
      }_{\Register \in \RegisterSet}
    \end{bmatrix}
    \FORALL \NodeVar[1]
    \Bigl(
      \smashoperator[r]{
        \bigOR_{
          \swl{
            \State[1] \in \HaltingStateSet:\, \OutputFunc(\State[1]) = \OutputYes
          }{
            \State[1] \in \HaltingStateSet
          }
        }
      }
      \NodeVar[1] \IN \SetVar[1]_{\State[1]} \,
    \Bigr),
  \end{equation*}
  which states that
  all \kl{nodes} end up in a \kl{halting state} that outputs~$\OutputYes$.

  Basically,
  the \kl{subformulas}~$\tuple{\Formula[1]_{\State[2]}}_{\State[2] \in \StateSet}$
  and~$\tuple{\Formula[1]_{\Register}}_{\Register \in \RegisterSet}$
  can be constructed in such a way that
  for all $i \in \Naturals$,
  the $(i + 1)$-th \kl{stage} of the \kl{partial fixpoint} induction
  represents the \kl{configuration}
  reached by~$\Automaton$ in the $i$-th round.
  To achieve this,
  each of the \kl{subformulas} contains a nested
  \kl[partial fixpoint operator]{partial fixpoint} \kl{formula}
  describing the result computed by the \kl{transition maker}~$\TransMaker$
  between two consecutive synchronous rounds,
  using additional \kl[set variables]{set} and \kl{function variables}
  to encode the \kl{inner configurations} of~$\TransMaker$ at each \kl{node}.
  Thus,
  each \kl{stage} of the nested \kl{partial fixpoint} induction
  corresponds to a single step
  in the \kl{transition maker}'s sequential scanning process.
  \qed
\end{proof}

Let us now consider the opposite direction
and sketch how to go from \kl{functional fixpoint logic}
to \kl{distributed register automata}
(see Appendix~\ref{app:logic-to-automata} for a full proof).

\begin{proposition}
  \label{prp:logic-to-automata}
  For every \kl{formula} of \kl{functional fixpoint logic}
  that \kl{defines} a \kl{graph property},
  we can construct
  an \kl{equivalent} \kl{distributed register automaton}.
\end{proposition}

\begin{proof}[sketch]
  We proceed by structural induction:
  each \kl{subformula}~$\Formula$
  will be evaluated by
  a dedicated \kl{automaton}~$\Automaton_{\Formula}$,
  and several such \kl{automata} can then be combined
  to build an \kl{automaton} for a composite \kl{formula}.
  For this purpose,
  it is convenient to design \emph{\kl{centralized}} \kl{automata},
  which operate on a given \kl{spanning tree}
  (as computed in Example~\ref{ex:spanning-tree})
  and are coordinated by the root
  in a fairly sequential manner.
  In~$\Automaton_{\Formula}$,
  each \kl{free} \kl{node variable}~$\NodeVar$ of~$\Formula$
  is represented by a corresponding \kl{input register}~$\NodeVar$
  whose value at the root is
  the current \kl{interpretation}~$\Interpret{\NodeVar}{\Assignment}$
  of~$\NodeVar$.
  Similarly,
  to represent a \kl{function variable}~$\FuncVar$,
  every \kl{node}~$\Node$ has a \kl{register}~$\FuncVar$
  storing~$\Interpret{\FuncVar}{\Assignment}(\Node)$.
  The \kl{nodes} also possess some \kl{auxiliary registers}
  whose purpose will be explained below.
  In the end,
  for any \kl{formula}~$\Formula$
  (potentially with \kl{free variables}),
  we will have an \kl{automaton}~$\Automaton_{\Formula}$
  computing a \kl{transduction}
  $\TransductionOf{\Automaton_{\Formula}} \colon
  \ConfigSet{{\InputLabelSet,\set{\Parent, \Root} \cup \free(\Formula)}} \to
  \ConfigSet{{\set{\OutputYes,\OutputNo},\EmptySet}}$,
  where
  $\Parent$ and $\Root$ are supposed to constitute a \kl{spanning tree}.
  The computation is triggered by the root,
  which means that
  the other \kl{nodes} are waiting for a signal to wake up.
  Essentially,
  the \kl{nodes} involved in the evaluation of~$\Formula$
  collect some information,
  send it towards the root,
  and go back to sleep.
  The root then returns~$\OutputYes$ or~$\OutputNo$,
  depending on whether or not~$\Formula$ holds
  in the \kl[input configuration]{input} \kl{graph}
  under the \kl{variable assignment} provided by the \kl{input registers}.
  Centralizing~$\Automaton_{\Formula}$ in that way
  makes it very convenient (albeit not efficient)
  to evaluate composite \kl{formulas}.
  For example,
  in~$\Automaton_{\Formula[1] \vee \Formula[2]}$,
  the root will first run~$\Automaton_{\Formula[1]}$,
  and then~$\Automaton_{\Formula[2]}$
  in case~$\Automaton_{\Formula[1]}$ returns~$\OutputNo$.

  \begin{algorithm}[t]
    \caption{\; $\Automaton_{\Formula}$ for
      $\Formula =
      \PFP [\FuncVar_i \DEF \Formula[1]_i]_{i \in \range[1]{\FuncNum}} \, \Formula[2]$,
      \,as controlled by the root}
    \label{algo:pfpmain}
    $\begin{array}{r@{\quad}l}
        1 & \Keyword{init}(\CounterIncAutomaton) \\[0.5ex]
        2 & \Keyword{repeat} \\
        3 & \qquad \Keyword{@every node do }
            \Keyword{for}~ i \in \range[1]{\FuncNum}
            ~\Keyword{do}~ \FuncVar_i \leftarrow \FuncVarNew_{i} \\
        4 & \qquad\Keyword{for}~ i \in \range[1]{\FuncNum}
            ~\Keyword{do}~\textit{update}(\FuncVarNew_{i}) \\
        5 & \qquad \Keyword{if}~
            \Keyword{@every node }(\forall i \in \range[1]{\FuncNum}:
            \FuncVarNew_{i} = \FuncVar_i)
            ~\Keyword{then}~ \Keyword{goto}~ 8 \\
        6 & \Keyword{until}~
            \Keyword{execute}(\CounterIncAutomaton) ~\Keyword{returns}~ \OutputNo
            \qquad
            \text{/$*$ until global counter at maximum $*$/} \\[1ex]
        7 & \Keyword{@every node do } \Keyword{for}~ i \in \range[1]{\FuncNum}
            ~\Keyword{do}~ \FuncVar_{i} \leftarrow \Self \\[0.5ex]
        8 & \Keyword{execute}(\Automaton_{\Formula[2]})
    \end{array}$
  \end{algorithm}

  The evaluation of atomic \kl{formulas} is straightforward.
  So let us focus on the most interesting case,
  namely when
  $\Formula =
  \PFP [\FuncVar_i \DEF \Formula[1]_i]_{i \in \range*{\FuncNum}} \, \Formula[2]$.
  The root's program is outlined in Algorithm~\ref{algo:pfpmain}.
  Line~1 initializes a counter
  that ranges from~$0$ to $n^{\CounterConstant n} - 1$,
  where~$n$ is the number of \kl{nodes}
  in the \kl[input configuration]{input} \kl{graph}.
  This counter is distributed in the sense that
  every \kl{node} has some dedicated \kl{registers}
  that together store the current counter value.
  Every execution of~$\CounterIncAutomaton$
  will increment the counter by~$1$,
  or return~$\OutputNo$
  if its maximum value has been exceeded.
  Now,
  in each iteration of the loop starting at Line~2,
  all \kl{registers}~$\FuncVar_i$ and~$\FuncVarNew_i$
  are updated in such a way that
  they represent the current and next \kl{stage},
  respectively,
  of the \kl{partial fixpoint} induction.
  For the former,
  it suffices that every \kl{node} copies,
  for all~$i$,
  the contents of~$\FuncVarNew_{i}$ to~$\FuncVar_i$ (Line~3).
  To update~$\FuncVarNew_i$,
  Line~4 calls a subroutine $\textit{update}(\FuncVarNew_i)$
  whose effect is that
  $\FuncVarNew_{i} =
  \FuncOperator_{\Formula[1]_i}((\FuncVar_i)_{i \in \range*{\FuncNum}})$
  for all~$i$,
  where
  $\FuncOperator_{\Formula[1]_i} \colon
  (\NodeSet^\NodeSet)^\FuncNum \to \NodeSet^\NodeSet$
  is the operator defined in Section~\ref{sec:logic}.
  Line~5 checks whether we have reached a fixpoint:
  The root asks every \kl{node} to compare, for all $i$,
  its \kl{registers}~$\FuncVarNew_{i}$ and~$\FuncVar_i$.
  The corresponding truth value is propagated back to the root,
  where $\mathit{false}$ is given preference over $\mathit{true}$.
  If the result is $\mathit{true}$,
  we exit the loop and proceed with calling~$\Automaton_{\Formula[2]}$
  to evaluate~$\Formula[2]$ (Line~8).
  Otherwise,
  we try to increment the global counter
  by executing~$\CounterIncAutomaton$ (Line~6).
  If the latter returns~$\OutputNo$,
  the fixpoint computation is aborted
  because we know that it has reached a cycle.
  In accordance with the \kl{partial fixpoint} semantics,
  all \kl{nodes} then write their own \kl{identifier}
  to every \kl{register}~$\FuncVar_i$ (Line~7)
  before~$\Formula[2]$ is evaluated (Line~8).
  \qed
\end{proof}



\section{Conclusion}
\label{sec:conclusion}

This paper makes some progress in the development
of a descriptive distributed complexity theory
by establishing a logical characterization
of a wide class of network algorithms,
modeled as \kl{distributed register automata}.

In our translation
from \kl[functional fixpoint logic]{logic} to \kl{automata},
we did not pay much attention to algorithmic efficiency.
In particular,
we made extensive use of
\kl[centralized automaton]{centralized} subroutines
that are triggered and controlled by a leader process.
A natural question for future research is to identify cases
where we can understand a distributed architecture
as an opportunity that allows us to evaluate \kl{formulas} faster.
In other words,
is there an expressive fragment of \kl{functional fixpoint logic}
that gives rise to efficient distributed algorithms
in terms of running time?
What about the required number of messages?
We are then entering the field of automatic
\emph{synthesis of practical distributed algorithms}
from logical specifications.
This is a worthwhile task,
as it is often much easier to declare what should be done
than how it should be done
(cf.~Examples~\ref{ex:hamiltonian-cycle-automaton}
and~\ref{ex:hamiltonian-cycle-formula}).

As far as the authors are aware,
this area is still relatively unexplored.
However,
one noteworthy advance was made
by Grumbach and Wu in~\cite{GrumbachW09},
where they investigated distributed evaluation
of first-order \kl{formulas}
on bounded-degree graphs and planar graphs.
We hope to follow up on this in future work.



\subsubsection*{Acknowledgments.}
We thank Matthias Függer for helpful discussions.
Work supported by
\href{http://www.lsv.fr/\%7Ebouyer/equalis}
      {ERC \emph{EQualIS} (FP7-308087)}
and
\href{https://www.irif.fr/anr/fredda/index}
      {ANR \emph{FREDDA} (17-CE40-0013)}.


\bibliographystyle{plainurl}
\bibliography{references}

\newpage
\appendix
\setcounter{algorithm}{0}

\section{Spanning-tree automaton}
\label{app:spanning-tree}

\newcommand{\Propa}[1]{\ensuremath{\textup{P}_{#1}}}


\begin{algorithm}[t]
  \caption{\; \kl{Transition maker}
    of the \kl{automaton} from Example~\ref{ex:spanning-tree}}
  \label{algo:spanning-tree}
  $\left.\parbox{33em}{$\begin{array}{@{\quad}l}
      \Keyword{if}~ \,\exists ~\text{\kl{neighbor}}~ \Neighbor \;
      (\Neighbor.\Root < \My.\Root) \colon \\[0.5ex]
      \qquad \My.\CurrentState \gets 1;
      \quad  \My.\Parent \gets \Neighbor.\Self;
      \quad  \My.\Root \gets \Neighbor.\Root
  \end{array}$}\right\}$ \Rule{1} \\[1.5ex]
  $\left.\parbox{33em}{$\begin{array}{@{\quad}l}
      \Keyword{else if}~ \My.\CurrentState = 1 \\[-1ex]
      \phantom{\Keyword{else if}}
        \land\:
        \text{$\forall$ \kl{neighbor} $\Neighbor
          \begin{bmatrix*}[l]
            \, \Neighbor.\Root = \My.\Root \;\land {} \\
            (\Neighbor.\Parent \neq \My.\Self \,\lor\,
              \Neighbor.\CurrentState = 2) \:\!
          \end{bmatrix*}$:
        } \\[-0.5ex]
      \qquad  \My.\CurrentState \gets 2
  \end{array}$}\right\}$ \Rule{2} \\[1.5ex]
  $\left.\parbox{33em}{$\begin{array}{@{\quad}l}
      \Keyword{else if}~
      (\My.\CurrentState = 2 \,\land\, \My.\Root = \My.\Self) \,\lor\,
      (\My.\Parent.\CurrentState = 3) \colon \\[0.5ex]
      \qquad \My.\CurrentState \gets 3
  \end{array}$}\right\}$ \Rule{3} \\[1.5ex]
  $\begin{array}{@{\quad}l}
      \Keyword{else}~ \,\text{do nothing}
  \end{array}$
\end{algorithm}


\subsection{Run of the spanning-tree automaton}

\begin{figure}[h]
  \centering
  \includegraphics[width=0.9\textwidth]{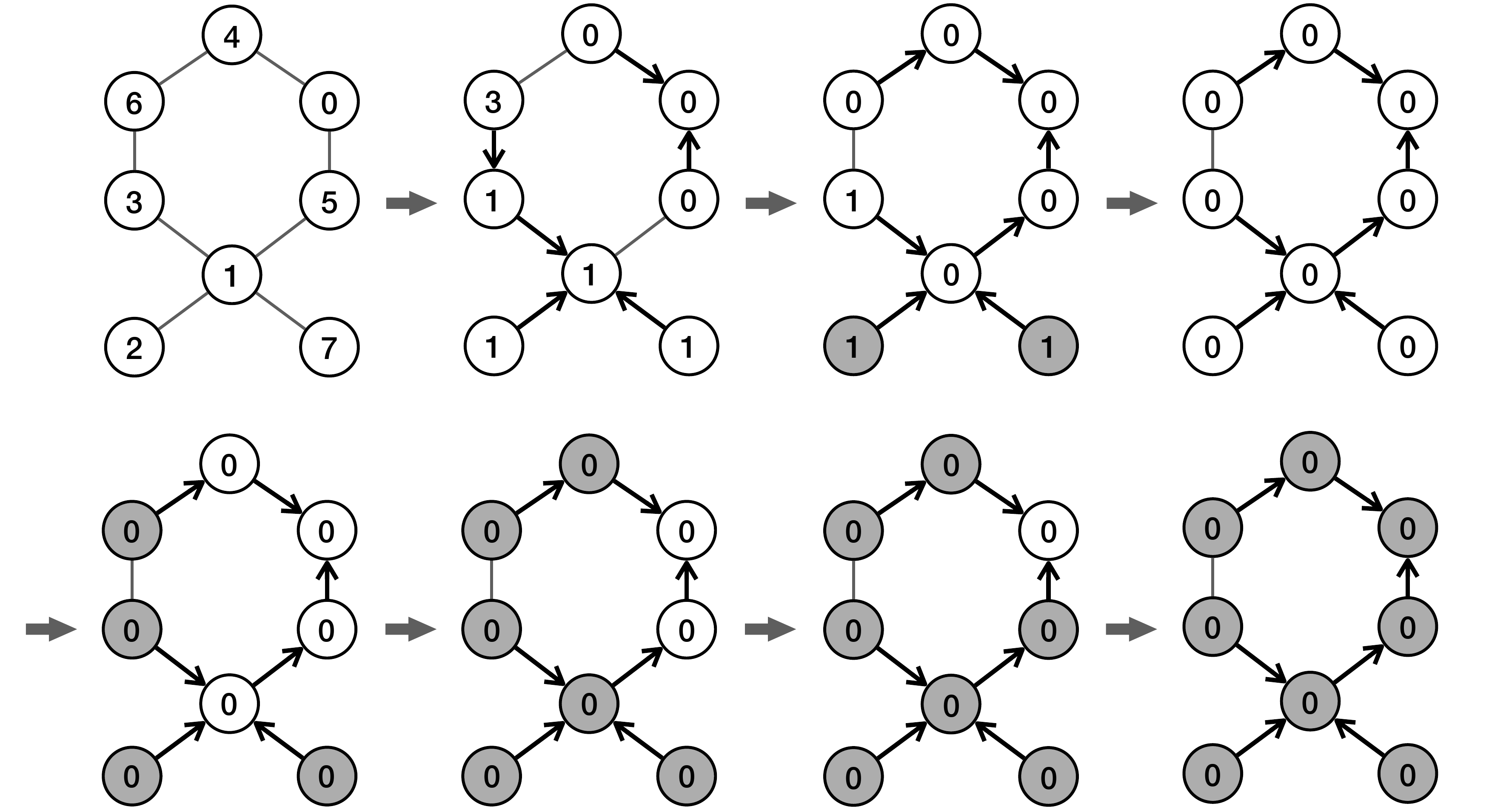} \\[2.3ex]
  \includegraphics[width=0.9\textwidth]{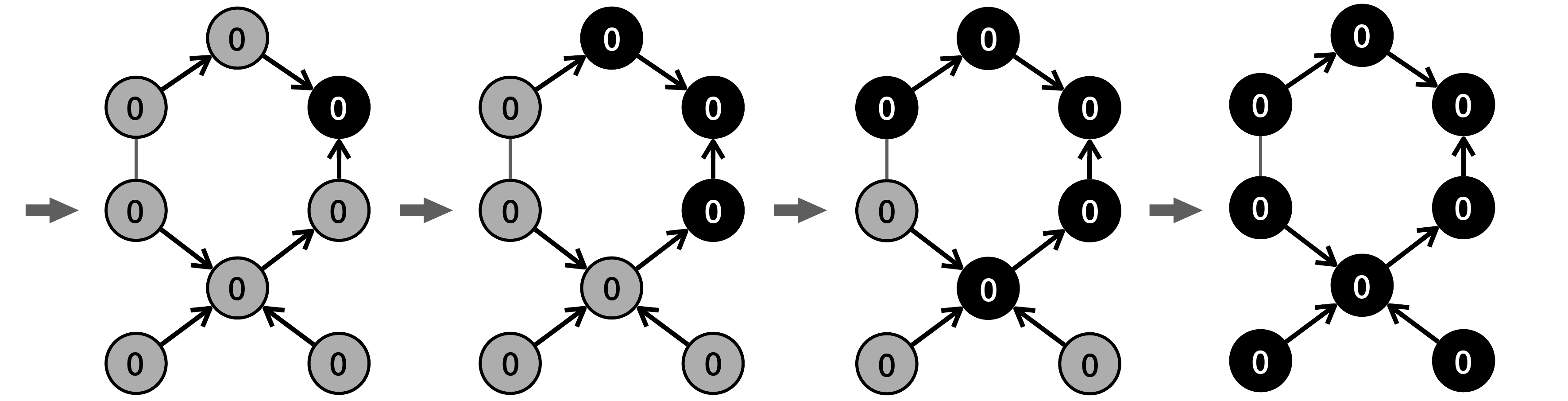}
  \caption{A run of the \kl{spanning-tree} \kl{automaton}
    from Example~\ref{ex:spanning-tree}}
  \label{fig:run-spanning-tree}
\end{figure}

We consider again the \kl{spanning-tree} \kl{automaton}
of Example~\ref{ex:spanning-tree} on page~\pageref{ex:spanning-tree}.
For convenience,
we recall its \kl{transition maker}
in Algorithm~\ref{algo:spanning-tree} above.
An example run is illustrated in Figure~\ref{fig:run-spanning-tree}.
The natural number within a \kl{node} is
the current content of its \kl{register}~$\Root$
(which, at the beginning of the run,
equals the \kl{identifier} of the respective \kl{node}).
The $\Parent$-relation is represented by thick arrows,
where we omit self loops.
Moreover,
white \kl{nodes} are in \kl{state}~$1$,
gray ones in \kl{state}~$2$,
and black ones in \kl{state}~$3$.
Note that \kl{nodes}~$2$ and~$7$ toggle between \kl{states}~$1$ and~$2$
before terminating in \kl{state}~$3$.
In the broadcast phase (last row),
\kl{node}~$0$ is the first one to enter \kl{state}~$3$.
The \kl{state} is then propagated to all other \kl{nodes}
to announce successful termination.

\subsection{Correctness of the spanning-tree automaton}

First,
we observe that in every \kl{graph}
$\Graph = \tuple{\NodeSet, \EdgeSet}$,
there must be a \kl{node}
that eventually enters \kl{state}~$3$.
Indeed,
it is straightforward to show by induction
that for every \kl{node}~$\Node[2]$ at \kl{distance}~$i$ of \kl{node}~$0$,
if~$\Node[2]$ has not reached \kl{state}~$3$ by time $i - 1$,
then for every time $t \geq i$,
we have $\Node[2].\Root = 0$
and $\Node[2].\Parent$ points to
some fixed \kl{node}~$\Node[2]'$
at \kl{distance} $i - 1$ of \kl{node}~$0$
(or \kl{distance}~$0$ in case $\Node[2] = 0$).
Note that~$\Node[2]$ will never modify its \kl{registers} again
once $\Node[2].\Root = 0$,
because the only rule that could potentially modify the \kl{registers},
i.e., \Rule{1},
is not applicable.
Hence,
if no \kl{node} has reached \kl{state}~$3$
after at most $\Diameter(\Graph)$ rounds of communication,
then the $\Parent$ pointers represent
a valid \kl{spanning tree} rooted at \kl{node}~$0$.
(Remember that~$\Graph$ is by definition connected.)
Since this tree remains forever unchanged,
it is easy to verify
that \kl{node}~$0$ will eventually enter \kl{state}~$3$,
which causes all other nodes to eventually do the same,
and thus the \kl{automaton} to \kl{halt}.
What remains to be shown is that no other \kl{node}
reaches \kl{state}~$3$ before \kl{node}~$0$ does.

To this end,
let us assume
that $\Node[1]$ is the first \kl{node} to enter \kl{state}~$3$,
and that it does so in the $(t+1)$-th round.
Therefore,
according to \Rule{3},
\begin{align}
  \text{at time $t$:} ~~
  \bigl(
    \Node[1].\CurrentState = 2 ~\wedge~
    \Node[1].\Root = \Node[1]
  \bigr) \,.
  \tag{a} \label{eq:stree}
\end{align}
Our goal is to show that~$\Node[1]$ is necessarily \kl{node}~$0$.

For $i \in \Naturals$,
let $\NodeSubset_i$ be the set of \kl{nodes}
at \kl{distance} exactly~$i$ from~$\Node[1]$.
Moreover,
for notational convenience,
we set $\NodeSubset_{-1} = \NodeSubset_0$
(which is~$\set{\Node[1]}$).
By a similar inductive argument as above,
we can easily see that
for all $i \in \Naturals$ and $\Node[2] \in \NodeSubset_{i+1}$,
\begin{align}
  \text{at every time $t' \in \Naturals$:} ~~
  \bigl(
    \Node[2].\Root = \Node[1] ~\implies~
    \Node[2].\Parent \in \NodeSubset_{i}
  \bigr) \,.
  \tag{b} \label{eq:shortest}
\end{align}

Based on~\eqref{eq:stree} and~\eqref{eq:shortest},
we now prove the following property:

\begin{lemma}
  \label{lemma:stree}
  For all $i \in \range{t}$, we have
  \begin{align}
    \text{at time $t-i$:} ~~~
    \text{$\forall \Node[2] \in \NodeSubset_i$:~ }
    \bigl(
      \Node[2].\CurrentState = 2 ~\wedge~
      \Node[2].\Root = \Node[1]
    \bigr) \,.
    \tag{\Propa{i}}
  \end{align}
\end{lemma}

\begin{proof}
  We proceed by induction.
  \Propa{0} is obvious,
  since it coincides with our assumption (\ref{eq:stree}).
  So suppose that \Propa{i} holds.
  We will show \Propa{i+1}.

  Take any \kl{node} $\Node[2] \in \NodeSubset_{i}$.
  By induction hypothesis,
  at time $t-i$,
  we have $\Node[2].\CurrentState = 2$
  and $\Node[2].\Root = \Node[1]$.
  Suppose that~$\Node[2]$ reaches such a \kl{local configuration}
  for the first time at time $t' \le t-i$.
  Note that~$\Node[2]$ cannot change its \kl{local configuration}
  between times~$t'$ and $t-i$
  (the only way to do so would be to choose another root,
  but then it could not return to root~$\Node[1]$ anymore).
  Let $\Node[2]' \in \NodeSubset_{i+1}$ be adjacent to~$\Node[2]$.
  According to \Rule{2},
  at time $t'-1$, we must have $\Node[2]'.\Root = \Node[1]$.
  Moreover,
  $\Node[2]'$ continues to have $\Node[2]'.\Root = \Node[1]$
  until time $t-(i+1)$
  (otherwise,
  $\Node[2]$ would have a different root than $\Node[1]$ at time $t-i$).

  Now,
  take any $\Node[2]' \in \NodeSubset_{i+1}$.
  We just showed that, at time $t-(i+1)$,
  we have $\Node[2]'.\Root = \Node[1]$.
  It remains to show that
  we also have $\Node[2]'.\CurrentState = 2$
  at time $t-(i+1)$.
  Let $\Node[2]$ be $\Node[2]'.\Parent$ at time $t-(i+1)$.
  By (\ref{eq:shortest}),
  we know that $\Node[2] \in \NodeSubset_i$.
  Let $t' \le t-i$ be as above,
  i.e., be the first point in time where
  $\Node[2].\CurrentState = 2$ and $\Node[2].\Root = \Node[1]$.
  According to \Rule{2},
  at time $t'-1$,
  we must have $\Node[2]'.\Root = \Node[1]$.
  As $\Node[2]'.\Root = \Node[1]$ at time $t-(i+1)$, too,
  $\Node[2]'.\Parent$ is the same for all
  $\hat{t} \in \range[t'-1]{t-(i+1)}$.
  In particular,
  $\Node[2]'.\Parent = \Node[2]$ at time $t'-1$.
  By \Rule{2},
  this implies
  $\Node[2]'.\CurrentState = 2$ at time $t'-1$,
  and therefore also
  $\Node[2]'.\CurrentState = 2$ at time $t-(i+1)$.
  This proves~\Propa{i+1}.
  \qed
\end{proof}

To conclude,
we argue that Lemma~\ref{lemma:stree} implies $\Node[1] = 0$.
It suffices to observe that for every \kl{node}~$\Node[2]$,
if $\Node[2].\Root = \Node[1]$ at any point in time,
then $\Node[1] \le \Node[2]$.
Since Lemma~\ref{lemma:stree} tells us that
every \kl{node}~$\Node[2]$ has~$\Node[1]$
in its \kl{register} $\Root$ at some point in time,
we can deduce that
$\Node[1] \le \Node[2]$
for every \kl{node}~$\Node[2] \in \NodeSet$,
which implies $\Node[1] = 0$.


\section{Syntactic sugar for functional fixpoint logic}
\label{app:syntactic-sugar}

\subsection{Encoding sets as functions}
\label{ssec:sets-as-functions}

To simplify the 
exposition and some of the 
proofs,
we have defined \kl{functional fixpoint logic}
in such a way that
the operator~$\PFP$ can bind only \kl{function variables}.
However,
since it is straightforward to encode sets of \kl{nodes} as functions,
we often take the liberty of writing \kl{formulas}
in which~$\PFP$ binds both
\kl[function variables]{function} and \kl{set variables}.
We now justify this formally
by showing how \kl{set variables} can always be eliminated.

\AP
To this end,
let us fix an infinite supply~$\SetVarSet$
of \kl{set variables}.
We extend the syntax of \kl{first-order logic}
to allow atomic \kl{formulas} of the form
$\Term \IN \SetVar$,
where~$\Term$ is a \kl{term}
and~$\SetVar$ is a \kl{set variable} in~$\SetVarSet$.
Naturally,
the semantics is that
“$\Term$~is an element of~$\SetVar$”,
and we will use
$\Term \NOTIN \SetVar$
as an abbreviation for
$\NOT (\Term \IN \SetVar)$.
The definitions of
\kl{free variables},
\kl{variable assignment},
\kl{interpretation}, and
\kl{satisfaction}
are generalized to \kl{set variables}
in the obvious way.

\AP
Since we consider only \kl{graphs}
that have at least two \kl{nodes}
and are equipped with a total order,
any set of \kl{nodes}~$\NodeSubset$
can be represented by a function
that maps
\kl{nodes} in~$\NodeSubset$ to their direct successors
and
\kl{nodes} outside of~$\NodeSubset$ to themselves.
The reason for choosing the direct successor
is simply because it is both easy to specify
and well-defined for all \kl{nodes} in all \kl{graphs},
assuming we consider the minimum \kl{node}
to be the direct successor of the maximum \kl{node}.
The following \kl{first-order} \kl{formula} schema states that
the \kl{node} represented by \kl{term}~$\Term[2]$ is the direct successor
of the \kl{node} represented by \kl{term}~$\Term[1]$:
\phantomintro{\FormulaSucc}
\begin{equation*}
  \reintro*\FormulaSucc[\Term[1], \Term[2]] \,\defeq\,
  \bigl(
    \Term[1] \SMALLER \Term[2] \,\AND
    \NOT \EXISTS \NodeVar[3] (
      \Term[1] \SMALLER \NodeVar[3] \,\AND\, \NodeVar[3] \SMALLER \Term[2])
  \bigr)
  \,\OR\:
  \FORALL \NodeVar[3] (
    \Term[2] \NOTGREATER \NodeVar[3] \,\AND\, \NodeVar[3] \NOTGREATER \Term[1])
\end{equation*}
(We write, for instance, “$\FormulaSucc[\NodeVar[1], \NodeVar[2]]$”
to instantiate this schema
with \kl{node variables}~$\NodeVar[1]$ and~$\NodeVar[2]$.)

Now,
let us consider a \kl{formula} of the form
$\PFP
 \bigl[
   \tuple{\FuncVar_i \DEF \Formula[1]_i}_{i \in \range*{\FuncNum}},
   \tuple{\SetVar \DEF \Formula[3]}
 \bigr] \,
 \Formula[2]$,
where
$\SetVar \in \SetVarSet$,\,
$\FuncVar_1, \dots, \FuncVar_{\FuncNum} \in \FuncVarSet$,
and $\Formula[1]_1, \dots, \Formula[1]_{\FuncNum}, \Formula[3], \Formula[2]$
are \kl{formulas}.
We assume that
the \kl{variable}~$\OutVar$ does not occur \kl{freely} in~$\Formula[3]$.
On an intuitive level,
since~$\SetVar$ is a \kl{set variable},
the set membership defined by~$\Formula[3]$
is with respect to a single special \kl{variable},
namely~$\InVar$.
A \kl{node}~$\Node$ lies in the corresponding set
precisely if~$\Formula[3]$ is \kl{satisfied}
by \kl{interpreting}~$\InVar$ as~$\Node$.
The \kl{stages} of the \kl{partial fixpoint} induction
are computed as before,
the only novelty being
that~$\SetVar$ is initialized to the empty set
(while the \kl{function variables}
$\FuncVar_1, \dots, \FuncVar_{\FuncNum}$
are still initialized to the identity function).
If the sequence of \kl{stages} does not converge to a fixpoint,
the \kl{partial fixpoint} is the same as the initial \kl{stage}
(i.e,~$\EmptySet$~for \kl{set variables}
and~$\Identity{\NodeSet}$ for \kl{function variables}).

We can easily eliminate~$\SetVar$ by replacing it with
a fresh \kl{function variable}~$\FuncVar_{\SetVar}$.
More precisely,
we rewrite the preceding \kl{formula} as
$\PFP
 \bigl[
   \tuple{\FuncVar_i \DEF \Formula[1]_i'}_{i \in \range*{\FuncNum}},
   \tuple{\FuncVar_{\SetVar} \DEF \Formula[3]'}
 \bigr] \,
 \Formula[2]'$.
To transform
$\tuple{\Formula[1]_i}_{i \in \range*{\FuncNum}}$ and~$\Formula[2]$
into their new forms
$\tuple{\Formula[1]_i'}_{i \in \range*{\FuncNum}}$ and~$\Formula[2]'$,
it suffices to replace every occurrence of an atomic \kl{formula}
of the form $\Term \IN \SetVar$
with its encoded representation
$\FuncVar_{\SetVar}(\Term) \NOTEQUAL \Term$.
That is, for instance,
$\Formula[2]' \defeq
 \Formula[2]
 \bigl[
   \Term \IN \SetVar \,\mapsto\, \FuncVar_{\SetVar}(\Term) \NOTEQUAL \Term
 \bigr]$.
For the \kl{subformula}~$\Formula[3]'$,
we additionally have to ensure
that~$\FuncVar_{\SetVar}$ is \kl{interpreted} as a function
that maps a \kl{node} to its direct successor
if and only if
it is contained in the new \kl{interpretation} of~$\SetVar$.
Thus,
we define
\begin{equation*}
  \Formula[3]' \,\defeq\,
  \Formula[3]
  \bigl[
    \Term \IN \SetVar \,\mapsto\, \FuncVar_{\SetVar}(\Term) \NOTEQUAL \Term
  \bigr]
  \,\AND\, \FormulaSucc[\InVar, \OutVar].
\end{equation*}

Notice that our encoding scheme preserves
the \kl{partial fixpoint} of the original \kl{formula}.
In particular,
if the sequence of \kl{stages} does not converge,
then~$\FuncVar_{\SetVar}$ is \kl{interpreted} as~$\Identity{\NodeSet}$,
which represents the empty set.

\subsection{Quantification over functions}
\label{ssec:function-quantifiers}

Since \kl{partial fixpoint} inductions allow us
to iterate over various \kl{interpretations}
of a \kl{function variable},
they provide a way of expressing
(second-order) quantification over functions.
To make this more convenient,
we add function quantifiers as “syntactic sugar”
to the language
of \kl{functional fixpoint logic}
and show how they can be converted into~$\PFP$ operators.

\AP
Consider a \kl{formula} of the form
$\EXISTSFUNC \FuncVar \, \Formula[1]$,
where~$\FuncVar$ is a \kl{function variable}
and~$\Formula[1]$ is a \kl{formula}.
Obviously,
the semantics is that
“there exists a function~$\FuncVar$ such that~$\Formula[1]$ holds”.
In the following,
we simulate this existential quantification
by a \kl{partial fixpoint} induction
that iterates over all possible \kl{interpretations} of~$\FuncVar$.
If there exists no \kl{interpretation} satisfying~$\Formula[1]$,
then the sequence of \kl{stages} of our induction
cycles forever through all functions,
never reaching a fixpoint.
Otherwise,
it ends up looping on
the first function that satisfies~$\Formula[1]$.
This function therefore becomes the \kl{interpretation}
assigned to~$\FuncVar$ by the \kl{partial fixpoint operator}.

To perform the iteration described above,
we need to define a cyclic order on the set of all functions.
This is easy because we restrict ourselves to
\kl{graphs} whose \kl{nodes} are totally ordered.
Thus,
each function on a given input \kl{graph}
$\Graph = \tuple{\NodeSet, \EdgeSet}$
can be thought of as
a $\card{\NodeSet}$-digit number written in base~$\card{\NodeSet}$,
where
the $i$-th least significant digit represents
the value taken by the function at the $i$-th smallest \kl{node}.
Based on this,
the direct successor of a function is simply
the function that corresponds to its number incremented by~$1$.
Additionally,
to make the order cyclic,
we stipulate that
the smallest function is the direct successor of the largest function;
in other words, there is an “integer overflow”.

It only remains to implement these ideas
in \kl{functional fixpoint logic}.
Without loss of generality,
we may assume that~$\Formula[1]$
does not contain any \kl{free} occurrences
of~$\InVar$ and~$\OutVar$.%
\footnote{%
  Otherwise,
  we replace
  $\EXISTSFUNC \FuncVar \, \Formula[1]$
  with the equivalent \kl{formula}
  $\EXISTS \NodeVar[1], \NodeVar[2]
   (\NodeVar[1] \EQUAL \InVar \,\AND\,
    \NodeVar[2] \EQUAL \OutVar \,\AND\,
    \EXISTSFUNC \FuncVar \,
    \Formula[1][\InVar,\OutVar \mapsto \NodeVar[1],\NodeVar[2]])$,
  where~$\Formula[1]$'s \kl{free} occurrences
  of~$\InVar$ and~$\OutVar$
  are substituted
  with fresh \kl{node variables}~$\NodeVar[1]$ and~$\NodeVar[2]$.
}
Hence,
we can rewrite
$\EXISTSFUNC \FuncVar \, \Formula[1]$
as the \kl{partial fixpoint} \kl{formula}\:\!
$\PFP[\FuncVar \DEF \Formula[2]] \, \Formula[1]$,
where
\begin{equation*}
  \Formula[2] \defeq\,
  \bigAND
  \begin{bmatrix*}[l]
    \phantom{\NOT}
    \bigl[
      \Formula[1] \,\OR\,
      \EXISTS \NodeVar[1], \NodeVar[2]
      \bigl(
        \NodeVar[1] \SMALLER \InVar \,\AND\,
        \FuncVar(\NodeVar[1]) \SMALLER \NodeVar[2]
      \bigr)
    \bigr]
    \;\IMP\:\,
    \OutVar \EQUAL \FuncVar(\InVar) \\[1ex]
    \NOT
    \bigl[
      \Formula[1] \,\OR\,
      \EXISTS \NodeVar[1], \NodeVar[2]
      \bigl(
        \NodeVar[1] \SMALLER \InVar \,\AND\,
        \FuncVar(\NodeVar[1]) \SMALLER \NodeVar[2]
      \bigr)
    \bigr]
    \;\IMP\:\,
    \FormulaSucc[\FuncVar(\InVar), \OutVar]
  \end{bmatrix*}.
\end{equation*}
The \kl{subformula}~$\Formula[2]$
distinguishes between three cases
in order to determine
how to update the \kl{interpretation} of~$\FuncVar$.
First,
as stated in the first line of the big conjunction,
if the current \kl{interpretation} already \kl{satisfies}~$\Formula[1]$,
then it is maintained.
Intuitively,
this can be read as:
“the new value~$\OutVar$ of~$\FuncVar$ at~$\InVar$
is equal to its current value~$\FuncVar(\InVar)$”.
Second,
even if~$\FuncVar$
(or rather the number representing it)
has to be incremented,
the function may retain its current value at some \kl{nodes}.
In fact,
“the value of digit~$\InVar$ remains unchanged
as long as there is some less significant digit~$\NodeVar[1]$
that can still be incremented”.
Third,
as stated in the second line of the big conjunction,
if none of the two previous cases apply,
then
“the value of digit~$\InVar$ must be
incremented if possible and otherwise reset”.
We express this using
the \kl{formula} schema~$\FormulaSucc[\Term[1], \Term[2]]$
from Section~\ref{ssec:sets-as-functions}
to ensure that
“the new value~$\OutVar$ is the direct successor
of the current value~$\FuncVar(\InVar)$.”

Now,
when we evaluate the \kl{formula}\:\!
$\PFP[\FuncVar \DEF \Formula[2]] \, \Formula[1]$\:\!
on a structure $\tuple{\Graph, \StateFunc}, \Assignment$,
there are two possibilities.
Either there exists no \kl{interpretation} of~$\FuncVar$
that \kl{satisfies}~$\Formula[1]$,
in which case
the \kl{partial fixpoint} induction does not reach a fixpoint.
This means that~$\FuncVar$
defaults to~$\Identity{\NodeSet}$
and\:\!
$\PFP[\FuncVar \DEF \Formula[2]] \, \Formula[1]$\:\!
evaluates to false.
Or~$\Formula[1]$ can be \kl{satisfied},
in which case the induction reaches a satisfying fixpoint.
This fixpoint is chosen as the \kl{interpretation} of~$\FuncVar$,
and\:\!
$\PFP[\FuncVar \DEF \Formula[2]] \, \Formula[1]$\:\!
evaluates to true.
In both cases,
the result is the same as when evaluating
$\EXISTSFUNC \FuncVar \, \Formula[1]$
on $\tuple{\Graph, \StateFunc}, \Assignment$.



\section{From automata to logic}
\label{app:automata-to-logic}

We now prove the first part of Theorem~\ref{thm:main}
(stated in Section~\ref{sec:introduction}),
by showing that
\kl{functional fixpoint logic} is at least as expressive
as \kl{distributed register automata}:

\begin{uProposition}[\ref{prp:automata-to-logic}]
  For every \kl{distributed register automaton}
  that \kl{decides} a \kl{graph property},
  we can effectively construct
  an \kl{equivalent} \kl{formula} of \kl{functional fixpoint logic}.
\end{uProposition}

The rest of this section is devoted to the proof of
Proposition~\ref{prp:automata-to-logic}.
We consider
a \kl{distributed register automaton}
$\Automaton =
 \tuple{\StateSet, \RegisterSet, \InputFunc, \TransMaker, \HaltingStateSet, \OutputFunc}$
with \kl{transition maker}
$\TransMaker = \tuple{\InnerStateSet, \InnerRegisterSet, \InnerInitState, \InnerTransFunc, \InnerOutputFunc}$
and assume
that~$\Automaton$ \kl{decides} a \kl{graph property}~$\Property$
over \kl{label} set~$\InputLabelSet$.
In the remainder of this section,
we construct a \kl{formula}~$\Formula[1]_{\Automaton}$
of \kl{functional fixpoint logic}
that \kl{defines}~$\Property$.
As our exposition goes into full detail,
it is a bit lengthy.
To make it clear how all the pieces fit together,
we present the construction in a top-down manner.

For each \kl{state}~$\State \in \StateSet$,
our \kl{formula} uses a \kl{set variable}~$\SetVar[1]_{\State}$
to represent the set of \kl{nodes} of the input \kl{graph}
that are in \kl{state}~$\State$.
Furthermore,
for each \kl{register}~$\Register \in \RegisterSet$,
it uses a \kl{function variable}~$\FuncVar[1]_{\Register}$
to represent the function that maps each \kl{node}~$\Node[1]$
to the \kl{node}~$\Node[2]$
whose \kl{identifier} is stored in~$\Node[1]$'s \kl{register}~$\Register$.
By means of a \kl{partial fixpoint operator},
we will ensure that
on any $\InputLabelSet$-\kl{labeled graph} $\tuple{\Graph, \StateFunc}$,
the \kl{interpretations}
of~$\tuple{\SetVar[1]_{\State}}_{\State \in \StateSet}$
and~$\tuple{\FuncVar[1]_{\Register}}_{\Register \in \RegisterSet}$
represent the \kl{halting configuration}
reached by~$\Automaton$ on~$\tuple{\Graph, \StateFunc}$.
Hence,
the final \kl{formula} is simply
\begin{equation*}
  \Formula[1]_{\Automaton} \defeq\:
  \PFP \!
  \begin{bmatrix}
    \tuple{\SetVar[1]_{\State[2]} \DEF \Formula[1]_{\State[2]}}_{\State[2] \in \StateSet} \\
    \tuple{\FuncVar[1]_{\Register} \DEF \Formula[1]_{\Register}}_{\Register \in \RegisterSet}
  \end{bmatrix}
  \FORALL \NodeVar[1]
  \Bigl(
    \smashoperator[r]{
      \bigOR_{
        \swl{
          \State[1] \in \HaltingStateSet:\, \OutputFunc(\State[1]) = \OutputYes
        }{
          \State[1] \in \HaltingStateSet
        }
      }
    }
    \NodeVar[1] \IN \SetVar[1]_{\State[1]} \,
  \Bigr),
\end{equation*}
which states that
all \kl{nodes} end up in a \kl{halting state} that outputs~$\OutputYes$.

\subsection{Simulating the automaton}

The real work is now to construct the
\kl{formulas}~$\tuple{\Formula[1]_{\State[2]}}_{\State[2] \in \StateSet}$
and~$\tuple{\Formula[1]_{\Register}}_{\Register \in \RegisterSet}$
in such a way that
for all $i \in \Naturals$,
the $(i + 1)$-th \kl{stage} of the \kl{partial fixpoint} induction
represents the \kl{configuration}
reached by~$\Automaton$ in the $i$-th round.
Note that in doing so,
we make sure that
the infinite sequence of \kl{stages} reaches a fixpoint
(given our assumption that
the \kl{automaton} is a \kl{decider} and thus eventually \kl{halts}).

For each \kl{state}~$\State[2] \in \StateSet$,
we update the \kl{set variable}~$\SetVar[1]_{\State[2]}$ with the \kl{formula}
\begin{equation*}
  \Formula[1]_{\State[2]} \defeq\,
  \Bigl(
    \smashoperator[r]{
      \bigOR_{
        \swl{
          \Label \in \InputLabelSet:\, \InputFunc(\Label) = \State[2]
        }{
          \Label \in \InputLabelSet
        }
      }
    }
    \LABELED{\Label} \InVar
    \,\AND\,
    \smashoperator{\bigAND_{\State[1] \in \StateSet}}
    \InVar \NOTIN \SetVar[1]_{\State[1]}
  \Bigr)
  \;\OR\;
  \Formula[1]_{\State[2]}^{\TransMaker}
  \;\OR\;
  \underbrace{
    \bigl( \InVar \IN \SetVar[1]_{\State[2]} \bigr).
  }_{\text{only if } \State[2] \in \HaltingStateSet}
\end{equation*}
Note that this \kl{formula} makes use of the syntactic sugar
introduced in Subsection~\ref{ssec:sets-as-functions},
which allows us to update \kl{set variables}
without any explicit reference to the \kl{variable}~$\OutVar$.
The first disjunct of~$\Formula[1]_{\State[2]}$ ensures that
\kl{stage}~$1$ of the \kl{partial fixpoint} induction
corresponds to the \kl{initial configuration} of the \kl{automaton}.
It does so by stating
that a \kl{node} (represented by the \kl{variable}~$\InVar$)
will belong to~$\SetVar[1]_{\State[2]}$ in the next \kl{stage}
if its \kl[input configuration]{input} \kl{label}~$\Label$
maps to~$\State[2]$
and it does not belong to any set~$\SetVar[1]_{\State[1]}$
in the current \kl{stage}.
The latter part holds only in \kl{stage}~$0$,
where all \kl{set variables} are initialized to~$\EmptySet$.
In the second disjunct,
we use the \kl{subformula}~$\Formula[1]_{\State[2]}^{\TransMaker}$
defined below
to ensure that if \kl{node}~$\InVar$ is currently active,
i.e., in a non-\kl{halting state},
then~$\InVar$ switches to \kl{state}~$\State[2]$
when executing the \kl{transition maker}~$\TransMaker$.
Finally,
in case~$\State[2]$ is a \kl{halting state},
the third disjunct of~$\Formula[1]_{\State[2]}$
states that~$\InVar$ remains in~$\State[2]$
if it is already there.
This formalizes the fact that \kl{halting states} are never left.

To implement
the above-mentioned \kl{subformula}~$\Formula[1]_{\State[2]}^{\TransMaker}$,
we use another \kl{partial fixpoint} induction,
which simulates
the behavior of the \kl{transition maker}~$\TransMaker$
between two consecutive synchronous rounds.
This second induction is thus nested
within the main induction of~$\Formula[1]_{\Automaton}$.
Similarly to before,
we introduce
a \kl{set variable}~$\SetVar[2]_{\InnerState}$
for each \kl{inner state}~$\InnerState$
and a \kl{function variable}~$\FuncVar[2]_{\InnerRegister}$
for each \kl{inner register}~$\InnerRegister$.
These \kl{variables} serve to encode
the \kl{inner configurations}
of the \kl{transition maker} at each \kl{node},
in the same way as
the \kl{variables}~$\tuple{\SetVar[1]_{\State}}_{\State \in \StateSet}$
and~$\tuple{\FuncVar[1]_{\Register}}_{\Register \in \RegisterSet}$
encode the \kl{local configurations} of the \kl{automaton}.
Furthermore,
we introduce
a \kl{function variable}~$\FuncVar[2]_{\pointer}$
to represent
the \kl{transition maker}'s reading head at each \kl{node}
and a \kl{set variable}~$\SetVar[2]_{\End}$
to represent
the set of \kl{nodes} that have finished
scanning their neighborhood.
We will make sure that
once the nested \kl{partial fixpoint} has been reached,
the \kl{variables}~$\tuple{\SetVar[2]_{\InnerState}}_{\InnerState \in \InnerStateSet}$
and~$\tuple{\FuncVar[2]_{\InnerRegister}}_{\InnerRegister \in \InnerRegisterSet}$
represent the final \kl{inner configurations}
reached by the \kl{nodes}.
Relying on that,
we define
\begin{equation*}
  \Formula[1]_{\State[2]}^{\TransMaker} \defeq\,
  \Bigl(
    \smashoperator[r]{
      \bigOR_{
        \swl{
          \State[1] \in \StateSet \setminus \HaltingStateSet
        }{
          \State[1] \in \StateSet
        }
      }
    }
    \InVar \IN \SetVar[1]_{\State[1]}
  \Bigr)
  \,\AND\;
  \PFP \!
  \begin{bmatrix}
    \tuple{
      \SetVar[2]_{\InnerState} \DEF \Formula[2]_{\InnerState}
    }_{\InnerState \in \InnerStateSet} \\
    \tuple{
      \FuncVar[2]_{\InnerRegister} \DEF \Formula[2]_{\InnerRegister}
    }_{\InnerRegister \in \InnerRegisterSet} \\
    \FuncVar[2]_{\pointer} \DEF \Formula[2]_{\pointer} \\
    \SetVar[2]_{\End} \DEF \Formula[2]_{\End}
  \end{bmatrix}
  \smashoperator[r]{
    \bigOR_{
      \substack{
        \InnerState \in \InnerStateSet \\[0.5ex]
        \swl{
          \exists \InnerOutputUpdateFunc:\,
          \InnerOutputFunc(\InnerState) = \tuple{\State[2], \InnerOutputUpdateFunc}
        }{
          \InnerState \in \InnerStateSet
        }
      }
    }
  }
  \InVar \IN \SetVar[2]_{\InnerState} \,.
\end{equation*}
The first conjunct simply checks
that~$\InVar$ is currently in a non-\kl{halting state},
which means that
it is allowed to run the \kl{transition maker}~$\TransMaker$.
The second conjunct expresses that
after running the \kl{transition maker},
$\InVar$ is in some \kl{inner state}~$\InnerState$
from which~$\TransMaker$ outputs \kl{state}~$\State[2]$.

Continuing with our top-down approach,
we first complete the discussion
of the outer \kl{partial fixpoint} induction
of~$\Formula[1]_{\Automaton}$
before turning to the nested one.
For each \kl{register}~$\Register \in \RegisterSet$,
the \kl{formula} with which we update~$\FuncVar[1]_{\Register}$ is
\begin{equation*}
  \Formula[1]_{\Register} \defeq\:
  \Formula[1]_{\Register}^{\TransMaker}
  \,\OR\,
  \Bigl(
    \smashoperator[r]{\bigOR_{\State[1] \in \HaltingStateSet}}
    \InVar \IN \SetVar[1]_{\State[1]}
    \,\AND\,
    \OutVar \EQUAL \FuncVar[1]_{\Register}(\InVar)
  \Bigr).
\end{equation*}
Here,
the \kl{subformula}~$\Formula[1]_{\Register}^{\TransMaker}$
states that~$\InVar$ is still in a non-\kl{halting state}
and writes the \kl{identifier} of~$\OutVar$
to its \kl{register}~$\Register$
after running the \kl{transition maker}.
The second disjunct of~$\Formula[1]_{\Register}$
covers the case
where~$\InVar$ has already reached a \kl{halting state},
forcing it to maintain
its current \kl[register valuation]{valuation} of~$\Register$.
Note that neither of the two disjuncts is satisfied
if all \kl{set variables}
are \kl{interpreted} as the empty set,
as is the case in \kl{stage}~$0$
of the (outer) \kl{partial fixpoint} induction.
Therefore,
$\Formula[1]_{\Register}$ does not define a functional relationship
in \kl{stage}~$0$,
which means that
the \kl{interpretation} of~$\FuncVar[1]_{\Register}$ in \kl{stage}~$1$
defaults to the identity function.
Rather conveniently,
this is precisely what we want,
i.e., \kl{stage}~$1$ represents
the \kl{initial configuration} of the \kl{automaton}.
Here we rely on the assumption
that~$\Automaton$ has no \kl{input registers}
(because it \kl{decides} a \kl{graph property})
and therefore initializes the \kl{registers} of each \kl{node}
to the \kl{node}'s own \kl{identifier}.

The implementation of~$\Formula[1]_{\Register}^{\TransMaker}$
is very similar to that of~$\Formula[1]_{\State[2]}^{\TransMaker}$.
In particular,
it uses the same nested induction to simulate~$\TransMaker$:
\begin{equation*}
  \Formula[1]_{\Register}^{\TransMaker} \defeq\,
  \Bigl(
    \smashoperator[r]{
      \bigOR_{
        \swl{
          \State[1] \in \StateSet \setminus \HaltingStateSet
        }{
          \State[1] \in \StateSet
        }
      }
    }
    \InVar \IN \SetVar[1]_{\State[1]}
  \Bigr)
  \,\AND\;
  \PFP \!
  \begin{bmatrix}
    \tuple{
      \SetVar[2]_{\InnerState} \DEF \Formula[2]_{\InnerState}
    }_{\InnerState \in \InnerStateSet} \\
    \tuple{
      \FuncVar[2]_{\InnerRegister} \DEF \Formula[2]_{\InnerRegister}
    }_{\InnerRegister \in \InnerRegisterSet} \\
    \FuncVar[2]_{\pointer} \DEF \Formula[2]_{\pointer} \\
    \SetVar[2]_{\End} \DEF \Formula[2]_{\End}
  \end{bmatrix}
  \smashoperator[r]{
    \bigOR_{
      \substack{
        \swl{
          \InnerState \in \InnerStateSet,\:\! \InnerRegister \in \InnerRegisterSet
        }{
          \InnerState \in \InnerStateSet
        } \\[0.5ex]
        \swl{
          \exists \State, \InnerOutputUpdateFunc:\,
          \InnerOutputFunc(\InnerState) = \tuple{\State, \InnerOutputUpdateFunc}
          \,\land\,
          \InnerOutputUpdateFunc(\Register) = \InnerRegister
        }{
          \InnerState \in \InnerStateSet
        }
      }
    }
  } \,
  \bigl(
    \InVar \IN \SetVar[2]_{\InnerState} \,\AND\,
    \OutVar \EQUAL \FuncVar[2]_{\InnerRegister}(\InVar)
  \bigr)
\end{equation*}
The only difference to~$\Formula[1]_{\State[2]}^{\TransMaker}$
is the big disjunction
within the scope of the~$\PFP$ operator.
It stipulates
that~$\InVar$ ends up in some \kl{inner state}~$\InnerState$
that causes the \kl{transition maker}
to update~$\InVar$'s \kl{register}~$\Register$
to the \kl{identifier} of~$\OutVar$.
For this to be true,
the \kl{identifier} of~$\OutVar$ must be stored in
the \kl{inner register}~$\InnerRegister$
that is used to update~$\Register$.

\subsection{Simulating the transition maker}

We now come to the nested \kl{partial fixpoint} induction.
As briefly mentioned above,
it uses two helper \kl{variables}~$\FuncVar[2]_{\pointer}$
and~$\SetVar[2]_{\End}$
to keep track of the \kl{transition maker}'s scanning process.
In each \kl{stage} of the induction,
$\FuncVar[2]_{\pointer}$ is \kl{interpreted}
as a function that maps each \kl{node}~$\Node[1]$
to the \kl{node}~$\Node[2]$
that is currently scanned
by (the \kl{transition maker} at)~$\Node[1]$.
In other words,
this function gives us
the current position of each \kl{node}'s reading head.
Since every \kl{node} starts
by reading its own \kl{local configuration},
it comes in handy that~$\FuncVar[2]_{\pointer}$
is initialized to the identity function at \kl{stage}~$0$
of the nested induction.
The function is then updated with the following \kl{formula}:
\begin{align*}
  \Formula[2]_{\pointer} \defeq\;\;
  & \InVar \LINKED \OutVar \;\AND {} \\
  & \hspace{-1ex}
    \bigOR
    \begin{bmatrix*}[l]
      \FuncVar[2]_{\pointer}(\InVar) \EQUAL \InVar \;\AND\;
      \FORALL \NodeVar
      \bigl(
        \InVar \LINKED \NodeVar
        \,\IMP\, \OutVar \NOTGREATER \NodeVar
      \bigr) \\[0.5ex]
      \FuncVar[2]_{\pointer}(\InVar) \NOTEQUAL \InVar \;\AND\;
      \FuncVar[2]_{\pointer}(\InVar) \SMALLER \OutVar \;\AND\;
      \FORALL \NodeVar
      \bigl(
        \InVar \LINKED \NodeVar \AND
        \FuncVar[2]_{\pointer}(\InVar) \SMALLER \NodeVar
        \,\IMP\, \OutVar \NOTGREATER \NodeVar
      \bigr) \\[0.5ex]
      \FuncVar[2]_{\pointer}(\InVar) \NOTEQUAL \InVar \;\AND\;
      \FuncVar[2]_{\pointer}(\InVar) \EQUAL \OutVar \;\AND\;
      \FORALL \NodeVar
      \bigl(
        \InVar \LINKED \NodeVar
        \,\IMP\, \NodeVar \NOTGREATER \OutVar
      \bigr)
    \end{bmatrix*}
\end{align*}
Here,
the first conjunct ensures that the reading head of~$\InVar$
can be moved only to a \kl{neighbor} of~$\InVar$,
while the big disjunction below
is responsible for selecting
the smallest \kl{neighbor} that has not yet been visited.
The first line of the disjunction covers the initial step,
where the reading head is still at~$\InVar$
and must be moved to the smallest of all \kl{neighbors}.
The second line corresponds to the case
where the head is moved
from one \kl{neighbor} to the next-smallest one.
Finally,
the third line states that
once the head has reached the greatest \kl{neighbor},
it remains there.
This is important to ensure that
the sequence of \kl{stages} converges to a fixpoint.
Also note that~$\Formula[2]_{\pointer}$
always defines a total function
because every \kl{node} has at least one \kl{neighbor}.
(This follows from our restriction to
connected \kl{graphs} with at least two \kl{nodes}.)

Since the reading head of the \kl{transition maker}
remains at the last-visited \kl{node},
we must prevent that \kl{node}
from being processed more than once.
This is the purpose of the \kl{set variable}~$\SetVar[2]_{\End}$,
which will be \kl{interpreted} as the set of all \kl{nodes}
that have finished scanning their neighborhood.
If~$\InVar$'s reading head
reaches the last \kl{neighbor} in \kl{stage}~$i$,
then~$\InVar$ is added to~$\SetVar[2]_{\End}$ in \kl{stage}~$i + 1$:
\begin{equation*}
  \Formula[2]_{\End} \defeq\:
  \FuncVar[2]_{\pointer}(\InVar) \NOTEQUAL \InVar \;\AND\;
  \FORALL \NodeVar
  \bigl(
    \InVar \LINKED \NodeVar
    \,\IMP\, \NodeVar \NOTGREATER \FuncVar[2]_{\pointer}(\InVar)
  \bigr)
\end{equation*}

Having formalized how
the \kl{transition maker}~$\TransMaker$ scans its input,
we now make use of
the \kl{variables}~$\FuncVar[2]_{\pointer}$ and~$\SetVar[2]_{\End}$
to define how it updates its \kl{inner configuration}.
For each \kl{inner state}~$\InnerState[2] \in \InnerStateSet$,
the \kl{set variable}~$ \SetVar[2]_{\InnerState[2]}$
is updated with the \kl{formula}
\begin{equation*}
  \Formula[2]_{\InnerState[2]} \defeq\,
  \Bigl(
    \InVar \IN \SetVar[2]_{\End} \,\AND\,
    \InVar \IN \SetVar[2]_{\InnerState[2]}
  \Bigr) \,\OR\,
  \Bigl(
    \InVar \NOTIN \SetVar[2]_{\End} \,\AND\,
    \smashoperator[r]{
      \bigOR_{
        \substack{
          \swl{
            \InnerState[1] \in \InnerStateSet,\,
            \State[1] \in \StateSet,\:\!
            {\SmallerRel} \in \powerset{(\InnerRegisterSet \cup \RegisterSet)^2}
          }{
            \InnerState[1] \in \InnerStateSet
          } \\[0.5ex]
          \swl{
            \exists \InnerUpdateFunc:\,
            \InnerTransFunc(\InnerState[1], \State[1], {\SmallerRel}) =
            \tuple{\InnerState[2], \InnerUpdateFunc}
          }{
            \InnerState[1] \in \InnerStateSet
          }
        }
      }
    }
    \Formula[3]_{(\InnerState[1], \State[1], {\SmallerRel})}^{\curr}
  \Bigr).
\end{equation*}
The first disjunct states
that~$\TransMaker$ remains in~$\InnerState[2]$
if it has reached the end of its input
and is currently in~$\InnerState[2]$,
whereas the second disjunct describes
an \kl{inner transition} to~$\InnerState[2]$
in case~$\TransMaker$ has not yet terminated.
For such an \kl{inner transition} to take place,
$\InnerState[2]$ must be the \kl{inner state}
that is obtained when applying~$\InnerTransFunc$ to
the current \kl{inner state}~$\InnerState[1]$ of~$\InVar$,
the current \kl{state}~$\State[1]$ of~$\FuncVar[2]_{\pointer}(\InVar)$,
and the relation~$\SmallerRel$
that compares the
\kl{inner register} values of~$\InVar$
with the
\kl{register} values of~$\FuncVar[2]_{\pointer}(\InVar)$.
Our \kl{formula} expresses this as a disjunction over
all possible choices of~$\InnerState[1]$,~$\State[1]$, and~$\SmallerRel$
that lead to~$\InnerState[2]$,
using
the \kl{subformula}~$\Formula[3]_{(\InnerState[1], \State[1], {\SmallerRel})}^{\curr}$
to check whether~$\InnerState[1]$,~$\State[1]$, and~$\SmallerRel$
do indeed correspond to the current
\kl{inner configuration} of~$\InVar$ and
\kl{local configuration} of~$\FuncVar[2]_{\pointer}(\InVar)$.

Implementing~$\Formula[3]_{(\InnerState[1], \State[1], {\SmallerRel})}^{\curr}$
for any
$\InnerState[1] \in \InnerStateSet$,\,
$\State[1] \in \StateSet$, and
${\SmallerRel} \in \powerset{(\InnerRegisterSet \cup \RegisterSet)^2}$
is straightforward:
\begin{align*}
  \Formula[3]_{(\InnerState[1], \State[1], {\SmallerRel})}^{\curr} \defeq\;\;
  &\Formula[3]_{\InnerState[1]}^{\curr} \;\AND\;
  \FuncVar[2]_{\pointer}(\InVar) \IN \SetVar[1]_{\State[1]} \;\AND {} \\[0.5ex]
  &\smashoperator{
    \bigAND_{
      \swl{
        \InnerRegister[2] \in \InnerRegisterSet,\:\!
        \Register[2] \in \RegisterSet :\,
        \InnerRegister[2] \SmallerRel \Register[2]
      }{
        \InnerRegister[2] \in \InnerRegisterSet
      }
    }
  } {}
  \FuncVar[2]_{\InnerRegister[2]}(\InVar)
  \SMALLER
  \FuncVar[1]_{\Register[2]}(\FuncVar[2]_{\pointer}(\InVar))
  \,\AND\,
  \smashoperator[r]{
    \bigAND_{
      \swl{
        \InnerRegister[2] \in \InnerRegisterSet,\:\!
        \Register[2] \in \RegisterSet :\,
        \Register[2] \SmallerRel \InnerRegister[2]
      }{
        \InnerRegister[2] \in \InnerRegisterSet
      }
    }
  } {}
  \FuncVar[1]_{\Register[2]}(\FuncVar[2]_{\pointer}(\InVar))
  \SMALLER
  \FuncVar[2]_{\InnerRegister[2]}(\InVar)
  \;\AND {} \\
  &\smashoperator{
    \bigAND_{
      \swl{
        \InnerRegister[2], \InnerRegister[2]' \in \InnerRegisterSet :\,
        \InnerRegister[2] \SmallerRel \InnerRegister[2]'
      }{
        \InnerRegister[2] \in \InnerRegisterSet
      }
    }
  } {}
  \swl{
    \FuncVar[2]_{\InnerRegister[2]}(\InVar)
    \SMALLER
    \FuncVar[2]_{\InnerRegister[2]'}(\InVar)
  }{
    \FuncVar[2]_{\InnerRegister[2]}(\InVar)
    \SMALLER
    \FuncVar[1]_{\Register[2]}(\FuncVar[2]_{\pointer}(\InVar))
  }
  \,\AND\,
  \smashoperator[r]{
    \bigAND_{
      \swl{
        \Register[2], \Register[2]' \in \RegisterSet :\,
        \Register[2] \SmallerRel \Register[2]'
      }{
        \Register[2] \in \RegisterSet
      }
    }
  } {}
  \FuncVar[1]_{\Register[2]}(\FuncVar[2]_{\pointer}(\InVar))
  \SMALLER
  \FuncVar[1]_{\Register[2]'}(\FuncVar[2]_{\pointer}(\InVar))
\end{align*}
In the first line,
we state
that~$\InnerState[1]$ is the \kl{inner state} of~$\InVar$
and~$\State[1]$ is the \kl{state} of~$\FuncVar[2]_{\pointer}(\InVar)$.
For the former,
we use the little helper \kl{formula}~$\Formula[3]_{\InnerState[1]}^{\curr}$
defined below.
In the remaining two lines,
we check that
all inequalities specified by~$\SmallerRel$ are satisfied.

The reason for using
the helper \kl{formula}~$\Formula[3]_{\InnerState[1]}^{\curr}$
is that the \kl{inner state} of~$\InVar$ is not represented explicitly
in \kl{stage}~$0$ of the nested induction.
Therefore,
if~$\InVar$ does not belong to any set~$\SetVar[2]_{\InnerState[2]}$,
we assume that
it is in the \kl{inner initial state}~$\InnerInitState$:
\begin{equation*}
  \Formula[3]_{\InnerState[1]}^{\curr} \defeq
  \begin{cases*}
    \InVar \IN \SetVar[2]_{\InnerState[1]} \,\OR\,
    \bigAND_{\InnerState[2] \in \InnerStateSet} \InVar \NOTIN \SetVar[2]_{\InnerState[2]}
    & if $\InnerState[1] = \InnerInitState$, \\
    \InVar \IN \SetVar[2]_{\InnerState[1]} & otherwise.
  \end{cases*}
\end{equation*}

Note that
since we represent each \kl{inner register}~$\InnerRegister$
by a \kl{function variable}~$\FuncVar[2]_{\InnerRegister}$,
our encoding does not take into account
that \kl{inner registers} can hold the undefined value~$\Undefined$.
In particular,
in \kl{stage}~$0$ of the nested induction,
$\FuncVar[2]_{\InnerRegister}$
is \kl{interpreted} as the identity function,
whereas the \kl{transition maker}
initializes~$\InnerRegister$ to~$\Undefined$.
However,
we may assume without loss of generality
that when the \kl{transition maker} starts
in its initial \kl{inner configuration}
$\tuple{
  \InnerInitState,
  \set{\InnerRegister \mapsto \Undefined}_{\InnerRegister \in \InnerRegisterSet}
}$
and reads the first \kl{local configuration}
$\tuple{\State[1], \RegisterVal}$
of its input sequence,
then it updates each of its \kl{inner registers}
to some value provided by~$\RegisterVal$
(and thus distinct from~$\Undefined$)
that depends only on~$\State[1]$ and the order relation
between the \kl{register} values of~$\RegisterVal$.
More formally,
we assume that
for every \kl{state} $\State[1] \in \StateSet$
and all binary relations
${\SmallerRel}, {\SmallerRel'} \subseteq (\InnerRegisterSet \cup \RegisterSet)^2$
such that
$\Register[1] \SmallerRel \Register[2]$
if and only if
$\Register[1] \SmallerRel' \Register[2]$
for all $\Register[1], \Register[2] \in \RegisterSet$,
we are guaranteed that
$\InnerTransFunc(\InnerInitState, \State[1], {\SmallerRel}) =
\InnerTransFunc(\InnerInitState, \State[1], {\SmallerRel'}) =
\tuple{\InnerState[2], \InnerUpdateFunc}$
such that
$\InnerUpdateFunc(\InnerRegister) \in \RegisterSet$
for all $\InnerRegister \in \InnerRegisterSet$.
On that basis,
we may substitute
the initial \kl{inner configuration} at \kl{node}~$\Node$
with the indistinguishable \kl{inner configuration}
$\tuple{
  \InnerInitState,
  \set{\InnerRegister \mapsto \Node}_{\InnerRegister \in \InnerRegisterSet}
}$,
which is precisely what we do in \kl{stage}~$0$.

To complete our construction,
it only remains to specify how the \kl{inner registers} are updated.
For each~$\InnerRegister[1] \in \InnerRegisterSet$,
we define the \kl{formula}
\begin{align*}
  \Formula[2]_{\InnerRegister[1]} \defeq\;\,
  &\bigl(
    \InVar \IN \SetVar[2]_{\End} \,\AND\,
    \OutVar \EQUAL \FuncVar[2]_{\InnerRegister[1]}(\InVar)
  \bigr) \;\OR {} \\
  &\biggl[
    \InVar \NOTIN \SetVar[2]_{\End} \,\AND\,
    \Bigl( \,
      \smashoperator{
        \bigOR_{
          \substack{
            \swl{
              \InnerState[1] \in \InnerStateSet,\,
              \State[1] \in \StateSet,\:\!
              {\SmallerRel} \in \powerset{(\InnerRegisterSet \cup \RegisterSet)^2}\!,\,
              \InnerRegister[2] \in \InnerRegisterSet
            }{
              \InnerState[1] \in \InnerStateSet
            } \\[0.5ex]
            \swl{
              \exists \InnerState[2], \InnerUpdateFunc:\,
              \InnerTransFunc(\InnerState[1], \State[1], {\SmallerRel}) =
              \tuple{\InnerState[2], \InnerUpdateFunc}
              \,\land\,
              \InnerUpdateFunc(\InnerRegister[1]) = \InnerRegister[2]
            }{
              \InnerState[1] \in \InnerStateSet
            }
          }
        }
      }
      \bigl[
        \Formula[3]_{(\InnerState[1], \State[1], {\SmallerRel})}^{\curr} \AND\,
        \OutVar \EQUAL \FuncVar[2]_{\InnerRegister[2]}(\InVar)
      \bigr]
      \,\OR
      \smashoperator[r]{
        \bigOR_{
          \substack{
            \swl{
              \InnerState[1] \in \InnerStateSet,\,
              \State[1] \in \StateSet,\:\!
              {\SmallerRel} \in \powerset{(\InnerRegisterSet \cup \RegisterSet)^2}\!,\,
              \Register[2] \in \RegisterSet
            }{
              \InnerState[1] \in \InnerStateSet
            } \\[0.5ex]
            \swl{
              \exists \InnerState[2], \InnerUpdateFunc:\,
              \InnerTransFunc(\InnerState[1], \State[1], {\SmallerRel}) =
              \tuple{\InnerState[2], \InnerUpdateFunc}
              \,\land\,
              \InnerUpdateFunc(\InnerRegister[1]) = \Register[2]
            }{
              \InnerState[1] \in \InnerStateSet
            }
          }
        }
      }
      \bigl[
        \Formula[3]_{(\InnerState[1], \State[1], {\SmallerRel})}^{\curr} \AND\,
        \OutVar \EQUAL \FuncVar[1]_{\Register[2]}(\FuncVar[2]_{\pointer}(\InVar))
      \bigr]
    \Bigr)
  \biggr],
\end{align*}
whose basic structure is very similar to that of~$\Formula[2]_{\InnerState[2]}$.
The first line just states that
the \kl{transition maker}~$\TransMaker$ retains
the current value of~$\InnerRegister[1]$
if it has reached the end of its input.
The second line covers the case
where~$\TransMaker$ has to
update~$\InVar$'s \kl{inner register}~$\InnerRegister[1]$
to a new value~$\OutVar$,
based on what it sees from
the current \kl{inner configuration} of~$\InVar$ and
the current \kl{local configuration} of~$\FuncVar[2]_{\pointer}(\InVar)$.
When~$\TransMaker$
evaluates its \kl{inner transition function}~$\InnerTransFunc$
on the currently seen
\kl{inner state}~$\InnerState[1]$,
\kl{state}~$\State[1]$,
and relation~$\SmallerRel$,
there are two possibilities:
either the new value of~$\InnerRegister[1]$
is obtained from
some \kl{inner register}~$\InnerRegister[2]$ of~$\InVar$,
or it is obtained from
some \kl{register}~$\Register[2]$ of~$\FuncVar[2]_{\pointer}(\InVar)$.
These two possibilities are expressed by
the two big disjunctions in the second line
of~$\Formula[2]_{\InnerRegister[1]}$.



\section{From logic to automata}
\label{app:logic-to-automata}

We will now take the opposite direction
and go from \kl{formulas} to \kl{automata}.
Hence,
this section is devoted to the proof of the following result:

\begin{uProposition}[\ref{prp:logic-to-automata}]
  For every \kl{formula} of \kl{functional fixpoint logic}
  that \kl{defines} a \kl{graph property},
  we can effectively construct
  an \kl{equivalent} \kl{distributed register automaton}.
\end{uProposition}

\AP
We proceed by induction,
i.e., we construct \kl{automata} for \kl{subformulas},
which will then be put together
to produce \kl{automata} for composed \kl{formulas}.
The invocation of \kl{automata} as subroutines will be more convenient
if they are \emph{\kl{centralized}}.
Intuitively,
this means that a dedicated root initiates an execution and,
at the end,
collects an acknowledgment from all the other processes before terminating.
Consider the set of \kl{registers}
$\intro*\TreeRegisterSet = \set{\Parent, \Root}$
and suppose
$\Config \in \ConfigSet{{\StateSet, \RegisterSet}}$
is a \kl{configuration} such that
$\TreeRegisterSet \subseteq \RegisterSet$.
We call~$\Config$ a \intro{spanning-tree configuration}
if the \kl[register valuations]{valuations} of~$\Parent$ and~$\Root$
form a \kl{spanning tree} in~$\Config$
as computed by the algorithm from Example~\ref{ex:spanning-tree}.

\begin{definition}[Centralized automaton]
  \label{def:centralized-automaton}
  A \intro{centralized automaton} is
  a \kl{distributed register automaton}~$\Automaton$
  with \kl{input registers}~$\InputRegisterSet$
  and \kl{output registers}~$\OutputRegisterSet$
  such that
  $\TreeRegisterSet \subseteq (\InputRegisterSet \setminus \OutputRegisterSet)$
  and all runs of~$\Automaton$ starting in
  a \kl[spanning-tree configuration]{spanning-tree} \kl{input configuration}
  satisfy the following properties:
  \begin{enumerate}
  \item \label{itm:initialization}
    A non-root \kl{node} can only leave
    its \kl[initial configuration]{initial} \kl{local configuration}
    after its parent changed its local \kl{state} for the first time.
  \item \label{itm:termination}
    The run eventually \kl{halts}
    and the root is the last \kl{node} to terminate
    (i.e., no \kl{node} enters a \kl{halting state} after the root does).
  \item \label{itm:register-preservation}
    The \kl{registers} in
    $(\InputRegisterSet \setminus \OutputRegisterSet)$
    are not modifed during the run.
  \end{enumerate}
\end{definition}

Essentially,
the notion of \kl{centralized automaton} provides a way
to combine and reason about \kl{distributed register automata}
in a fully sequential way.
Given a \kl{centralized automaton}
$\Automaton_1 =
\tuple{\StateSet_1, \RegisterSet_1, \InputFunc_1,
       \TransMaker_1, \HaltingStateSet_1, \OutputFunc_1}$
with \kl{input registers}~$\InputRegisterSet_1$
and \kl{output registers}~$\OutputRegisterSet_1$,
we can construct another \kl{centralized automaton}
$\Automaton_2 =
\tuple{\StateSet_2, \RegisterSet_2, \InputFunc_2,
       \TransMaker_2, \HaltingStateSet_2, \OutputFunc_2}$
that uses~$\Automaton_1$ as a subroutine.
To this end,
we make sure
that~$\StateSet_2$ is a Cartesian product of the form
${\ComponentSet \times \set{\Main, \Call} \times \StateSet_1}$
and that~$\RegisterSet_2$ includes~$\RegisterSet_1$.
In any execution of~$\Automaton_2$,
every \kl{node} is initially in the $\Main$ mode
(by which we mean that
the relevant component of its \kl{state} is $\Main$).
While in this mode,
a \kl{node} behaves according to
the algorithm implemented by~$\Automaton_2$.
The algorithm must be such that at some point in time,
all \kl{nodes} except the root of the \kl{spanning tree} are idle,
and the current \kl{configuration} of~$\Automaton_2$
yields a desired \kl{initial configuration} of~$\Automaton_1$
if we project all \kl{states} to their $\StateSet_1$-component
and consider only those \kl{registers} that belong to~$\RegisterSet_1$.
Then,
the root can launch a nested execution of~$\Automaton_1$
by entering the $\Call$ mode.
While in this mode,
a \kl{node} must simulate \kl{automaton}~$\Automaton_1$,
using only the $\StateSet_1$-component of its \kl{state}
and its $\RegisterSet_1$-\kl{registers}.
Whenever a \kl{node} in the $\Main$ mode sees that
one of its \kl{neighbors} has entered the $\Call$ mode,
it does the same.
Since~$\Automaton_1$ is \kl{centralized},
by Definition~\ref{def:centralized-automaton}\,(\ref{itm:initialization}),
this approach yields a faithful simulation of~$\Automaton_1$,
despite the fact that
the \kl{nodes} do not enter the $\Call$ mode simultaneously.
Moreover,
by Definition~\ref{def:centralized-automaton}\,(\ref{itm:termination}),
once the root has reached a \kl{halting state} of~$\Automaton_1$,
it knows that the \kl{automaton} has \kl{halted} everywhere.
The root can thus switch back to the $\Main$ mode
and resume executing~$\Automaton_2$.
The other \kl{nodes} do the same
as soon as one of their \kl{neighbors} has done so.
As a result,
\kl{automaton}~$\Automaton_2$ can use
the \kl{output configuration} produced by~$\Automaton_1$.
Since
by Definition~\ref{def:centralized-automaton}\,(\ref{itm:register-preservation}),
$\Automaton_1$ does not modify the \kl{registers} in
$\InputRegisterSet_1 \setminus \OutputRegisterSet_1$,
we can rely on them being the same as before calling~$\Automaton_1$.
In particular,
the \kl{spanning tree} represented by the \kl{registers} in~$\TreeRegisterSet$
remains unchanged.

Note that
while the preceding approach may lead to very inefficient algorithms
from the perspective of distributed computing,
it is sufficient for the purposes of this paper,
since we only want to characterize the fundamental expressive power
of \kl{distributed register automata}.

\smallskip

As our proof of Proposition~\ref{prp:logic-to-automata}
will proceed by induction on the structure of \kl{formulas},
we will have to cope with the \kl{interpretations}
of \kl{free} \kl[node variables]{node} and \kl{function variables}.
To this end,
we are going to encode \kl{interpretations} into \kl{configurations}.
Consider a \kl{spanning-tree configuration}
$\Config =
\tuple{\tuple{\NodeSet,\EdgeSet}, \StateFunc, \RegisterValFunc}
\in \ConfigSet{{\StateSet, \RegisterSet}}$,
with root $\Node_0$,
such that the set~$\RegisterSet$ contains \kl{variables} from
$\NodeVarSet \cup \FuncVarSet$.
Then,
$\Config$ defines a \kl{variable assignment}~$\Assignment$
as follows:
For a \kl{node variable} $\NodeVar \in \RegisterSet$,
we let
$\Interpret{\NodeVar}{\Assignment} = \RegisterValFunc(\Node_0)(\NodeVar)$.
That is,
the \kl{interpretation} of \kl{variable}~$\NodeVar$ is
the value of \kl{register}~$\NodeVar$ at the root.
Moreover,
each \kl{register}
$\FuncVar \in \RegisterSet \cap \FuncVarSet$
encodes an \kl{interpretation}
$\Interpret{\FuncVar}{\Assignment}$
of the \kl{function variable}~$\FuncVar$
by letting
$\Interpret{\FuncVar}{\Assignment}(\Node) =
\RegisterValFunc(\Node)(\FuncVar)$
for all $\Node \in \NodeSet$.

The next preparatory proposition states
that there is a \kl{centralized automaton}
that evaluates a \kl{term} with respect to
the \kl{variable assignment} encoded in the \kl{input configuration}.

\begin{proposition}[Automata for terms]
  \label{prop:Aterm}
  Let~$\Term$ be a \kl{term}.
  There is a \kl{centralized automaton}~$\Automaton_{\Term}$
  computing a \kl{transduction}
  \begin{equation*}
    \TransductionOf{\Automaton_{\Term}} \colon
    \ConfigSet{{\Singleton,\TreeRegisterSet \cup \free(\Term)}} \to
    \ConfigSet{{\Singleton,\set{\Term}}}
  \end{equation*}
  such that,
  given a
  \kl[spanning-tree configuration]{spanning-tree} \kl{input configuration}
  $\Config =
  \tuple{\Graph, \StateFunc, \RegisterValFunc}$
  with
  $\Graph = \tuple{\NodeSet, \EdgeSet}$
  and root $\Node_0 \in \NodeSet$
  encoding a \kl{variable assignment}~$\Assignment$,
  the \kl{automaton} eventually \kl[output configuration]{outputs}
  a \kl{configuration}
  $\Config' = \tuple{\Graph, \StateFunc, \RegisterValFunc'}$
  satisfying
  $\RegisterValFunc'(\Node_0)(\Term) = \Interpret{\Term}{\Assignment}$.
\end{proposition}

\begin{proof}
  Let
  $\Term= \FuncVar_{\FuncNum}(\ldots \FuncVar_2(\FuncVar_1(\NodeVar)) \ldots)$.
  Note that
  $\free(\Term)=\set{\FuncVar_1, \ldots, \FuncVar_{\FuncNum}, \NodeVar}$.
  The case $\FuncNum = 0$ is trivial,
  so suppose $\FuncNum \ge 1$.
  For $i \in \range[1]{\FuncNum}$,
  let
  $\Term_i = \FuncVar_i(\ldots\FuncVar_1(\NodeVar)\ldots)$.

  Given
  a \kl[spanning-tree configuration]{spanning-tree} \kl{input configuration}
  $\Config = \tuple{\tuple{\NodeSet,\EdgeSet}, \StateFunc, \RegisterValFunc}$,
  the \kl{automaton}~$\Automaton_{\Term}$ will use
  \kl{auxiliary registers}~$\Term_i$ to compute and store
  the intermediary values
  $\Interpret{\Term_i}{\Assignment} =
  \Interpret{\FuncVar_i}{\Assignment}
  (\ldots
    \Interpret{\FuncVar_1}{\Assignment}(\Interpret{\NodeVar}{\Assignment})
  \ldots)$.
  The \kl{register} $\Term = \Term_{\FuncNum}$
  is the only \kl{output register}.
  We compute the values for $\Term_1,\ldots,\Term_{\FuncNum}$ successively.
  As an invariant,
  we maintain that,
  after computing $\Interpret{\Term_i}{\Assignment}$,
  the root stores $\Interpret{\Term_i}{\Assignment}$
  in its \kl{register}~$\Term_i$.

  We reserve an \kl{auxiliary register}~$\Self$
  for storing the own \kl{identifier} of a \kl{node}.
  To compute $\Interpret{\Term_{i+1}}{\Assignment}$,
  we proceed as follows.
  Starting from the root,
  the value of~$\Interpret{\Term_i}{\Assignment}$
  is propagated to all \kl{nodes}
  along the \kl{spanning tree} down to the leaves.
  The leaves, in turn, initiate a back-propagation phase,
  where an internal \kl{node} waits for signals from all its children,
  notifies its parent,
  and then terminates.
  However,
  during this back-propagation,
  the \kl{node} whose \kl{identifier} coincides with~$\Term_i$
  includes the contents of its \kl{register}~$\FuncVar_{i+1}$
  in its message to its parent \kl{node},
  which propagates it further,
  until the value reaches the root.
  The latter stores it in its \kl{register}~$\Term_{i+1}$.
  \qed
\end{proof}

As further preparation for
the inductive translation of \kl{formulas},
we describe a subroutine that allows the root
to increment a \kl{register} by one.

\begin{proposition}[Register incrementation]
  \label{prop:inc}
  Let~$\Register$ be a \kl{register}.
  There is a \kl{centralized automaton}~$\RegIncAutomaton{\Register}$
  computing a \kl{transduction}
  \begin{equation*}
    \TransductionOf{\RegIncAutomaton{\Register}} \colon
    \ConfigSet{{\Singleton,\TreeRegisterSet \cup \set{\Register}}} \to
    \ConfigSet{{\Singleton, \set{\Register}}}
  \end{equation*}
  such that,
  given a
  \kl[spanning-tree configuration]{spanning-tree} \kl{input configuration}
  $\Config = \tuple{\Graph, \StateFunc, \RegisterValFunc}$
  with $\Graph = \tuple{\NodeSet,\EdgeSet}$
  and root $\Node_0 \in \NodeSet$,
  the \kl{automaton} eventually \kl[output configuration]{outputs}
  a \kl{configuration}
  $\Config' = \tuple{\Graph, \StateFunc, \RegisterValFunc'}$
  satisfying
  $\RegisterValFunc'(\Node_0)(\Register) =
  \min \set{\RegisterValFunc(\Node_0)(\Register) + 1,\, \card{\NodeSet} - 1}$.
\end{proposition}

\begin{proof}
  First,
  the current value~$\Node[3]$ of~$\Node_0$'s \kl{register}~$\Register$
  is broadcast through the entire \kl{graph}.
  Then,
  every \kl{node}~$\Node[1]$ sends to its parent in the \kl{spanning tree}
  the smallest value greater than~$\Node[3]$
  among the \kl{node} \kl{identifiers} in the subtree rooted at~$\Node[1]$
  (provided that this subtree contains such an \kl{identifier}).
  This procedure starts at the leaves and works its way up to the root.
  There,
  the smallest value obtained must be $\Node[3] + 1$.
  \qed
\end{proof}

We are now ready to transform any \kl{formula}
into an \kl{equivalent} \kl{automaton}.

\begin{proposition}
  \label{prop:logic2aut}
  Let~$\Formula$ be a \kl{formula} of \kl{functional fixpoint logic}
  over some finite set of \kl{labels}~$\InputLabelSet$.
  There is a \kl{centralized automaton}~$\Automaton_{\Formula}$
  computing a \kl{transduction}
  \begin{equation*}
    \TransductionOf{\Automaton_{\Formula}} \colon
    \ConfigSet{{\InputLabelSet,\TreeRegisterSet \cup \free(\Formula)}} \to
    \ConfigSet{{\set{\OutputYes,\OutputNo},\EmptySet}}
  \end{equation*}
  such that,
  given a
  \kl[spanning-tree configuration]{spanning-tree} \kl{input configuration}
  $\Config = \tuple{\Graph, \StateFunc, \RegisterValFunc}$
  with
  $\Graph = \tuple{\NodeSet,\EdgeSet}$
  and root $\Node_0$
  encoding a \kl{variable assignment}~$\Assignment$,
  the \kl{automaton} eventually \kl[output configuration]{outputs}
  a \kl{configuration}
  $\Config' = \tuple{\Graph, \StateFunc', \RegisterValFunc'}$
  satisfying
  $\StateFunc'(\Node_0) = \OutputYes$
  if and only if
  $\tuple{\Graph, \StateFunc}, \Assignment \,\SAT\, \Formula$.
\end{proposition}

\begin{proof}
  We proceed by induction on the structure of \kl{formulas}.

  \paragraph*{Case $\Formula = \LABELED{\Label} \Term$.}
  Given
  a \kl[spanning-tree configuration]{spanning-tree} \kl{input configuration}
  $\Config = \tuple{\tuple{\NodeSet,\EdgeSet}, \StateFunc, \RegisterValFunc}$
  encoding a \kl{variable assignment}~$\Assignment$,
  the \kl{automaton}~$\Automaton_{\Formula}$ first applies,
  as a subroutine,
  $\Automaton_{\Term}$ from Proposition~\ref{prop:Aterm}.
  The latter eventually \kl[output configuration]{outputs}
  a \kl{configuration} in
  $\ConfigSet{{\Singleton,\set{\Term}}}$
  such that
  the value of \kl{register}~$\Term$ at the root is
  $\Interpret{\Term}{\Assignment}$.
  Similarly to the construction of Proposition~\ref{prop:Aterm},
  the root then broadcasts~$\Interpret{\Term}{\Assignment}$
  to all other \kl{nodes} along the \kl{spanning tree},
  down to the leaves.
  During a subsequent back-propagation phase,
  \kl{node}~$\Interpret{\Term}{\Assignment}$ checks
  whether its \kl{label} is equal to~$\Label$.
  Accordingly,
  it sends either~$\OutputYes$ or~$\OutputNo$ to its parent,
  which propagates it further until it reaches the root.

  \paragraph*{Case $\Formula = \Term[1] \SMALLER \Term[2]$.}
  As a subroutine,
  the root first launches~$\Automaton_{\Term[1]}$
  and then~$\Automaton_{\Term[2]}$
  so that it eventually stores,
  in (\kl[auxiliary registers]{auxiliary})
  \kl{registers}~$\Term[1]$ and~$\Term[2]$,
  the \kl{interpretation}~$\Interpret{\Term[1]}{\Assignment}$
  and~$\Interpret{\Term[2]}{\Assignment}$,
  respectively.
  The root then compares both values
  and outputs the corresponding truth value.

  \paragraph*{Case $\Formula = \Term[1] \LINKED \Term[2]$.}
  Like in the previous case,
  the root first launches~$\Automaton_{\Term[1]}$
  and then~$\Automaton_{\Term[2]}$
  so that it eventually stores~$\Interpret{\Term[1]}{\Assignment}$
  and~$\Interpret{\Term[2]}{\Assignment}$
  in \kl{registers}~$\Term[1]$ and~$\Term[2]$,
  respectively.
  Both are then propagated along the \kl{spanning tree}.
  During a subsequent back-propagation,
  \kl{node}~$\Interpret{\Term[1]}{\Assignment}$ checks
  whether~$\Interpret{\Term[2]}{\Assignment}$ is in its neighborhood.
  The corresponding truth value is forwarded back to the root.

  \paragraph*{Case $\Formula = \NOT \Formula[2]$.}
  As a subroutine,
  $\Automaton_{\NOT\Formula[2]}$ first applies~$\Automaton_{\Formula[2]}$
  to the given
  \kl[spanning-tree configuration]{spanning-tree} \kl{input configuration}.
  Upon termination,
  the root will just flip the \kl{node} \kl{label}
  from~$\OutputYes$ to~$\OutputNo$ or vice versa.

  \paragraph*{Case $\Formula = \Formula_1 \OR \Formula_2$.}
  The \kl{automaton}~$\Automaton_{\Formula}$
  first applies~$\Automaton_{\Formula_1}$
  as a subroutine.
  If the root outputs~$\OutputYes$,
  then it stops.
  Otherwise,
  $\Automaton_{\Formula_2}$ is launched.

  \paragraph*{Case $\Formula = \EXISTS \NodeVar \, \Formula[2]$.}
  Automaton~$\Automaton_{\Formula}$
  has an \kl{auxiliary register}~$\NodeVar$.
  It makes use of the subroutine~$\RegIncAutomaton{\NodeVar}$
  from Proposition~\ref{prop:inc},
  which increments the value of~$\NodeVar$ by one
  every time it is called.
  The root,
  initially storing its own \kl{identifier}~$0$ in~$\NodeVar$,
  starts by launching~$\Automaton_{\Formula[2]}$.
  If the latter outputs~$\OutputYes$ (at the root),
  then the root outputs~$\OutputYes$ and stops.
  Otherwise,
  using~$\RegIncAutomaton{\NodeVar}$,
  the root will increment its \kl{register}~$\NodeVar$ by one,
  launch~$\Automaton_{\Formula[2]}$ again,
  and so on.
  If no increment is possible anymore,
  $\Automaton_{\Formula}$ outputs $\OutputNo$.

  \paragraph*{Case
    $\Formula =
    \PFP [\FuncVar_i \DEF \Formula[1]_i]_{i \in \range*{\FuncNum}} \, \Formula[2]$.}
  First of all,
  recall that $\free(\Formula)$ is the union of all sets
  $\free(\Formula[1]_i)
  \setminus \set{\FuncVar_1, \dots, \FuncVar_\FuncNum, \InVar, \OutVar}$
  and
  $\free(\Formula[2])
  \setminus \set{\FuncVar_1, \dots, \FuncVar_\FuncNum}$.
  We are interested in a \kl{transduction}
  $\TransductionOf{\Automaton_{\Formula}} \colon
  \ConfigSet{{\InputLabelSet, \TreeRegisterSet \cup \free(\Formula)}} \to
  \ConfigSet{{\set{\OutputYes, \OutputNo}, \EmptySet}}$.
  To implement the \kl{partial fixpoint} computation,
  we introduce \kl{auxiliary registers}
  $\FuncVar_1, \ldots, \FuncVar_\FuncNum$
  (for the current \kl{interpretation})
  and
  $\FuncVarNew_{1}, \ldots, \FuncVarNew_{\FuncNum}$
  (for the next \kl{interpretation}).
  We also use \kl{auxiliary registers}~$\InVarReg$ and~$\OutVarReg$,
  as well as a \kl{register}~$\Self$
  for storing the own \kl{identifier} of each \kl{node}.
  Furthermore,
  to test whether the \kl{partial fixpoint} computation
  has reached a cycle with a period greater than one,
  we include an additional set of \kl{auxiliary registers}
  that will allow us to implement a global counter.
  Since there are precisely~$n^{\FuncNum n}$
  different $\FuncNum$-tuples of functions
  on an \kl[input configuration]{input} \kl{graph}
  with~$n$ \kl{nodes},
  it is sufficient
  to count from~$0$ to $n^{\FuncNum n} - 1$.
  The counter is taken for granted for now,
  but it will be explained below,
  and we shall define a helper \kl{automaton}~$\CounterIncAutomaton$
  that allows us to increment it by one.

  \begin{algorithm}[t]
    \caption{\; $\Automaton_{\Formula}$ for
      $\Formula =
      \PFP [\FuncVar_i \DEF \Formula[1]_i]_{i \in \range[1]{\FuncNum}} \, \Formula[2]$,
      \,as controlled by the root}
    \label{algo:pfp}
    $\begin{array}{r@{\quad}l}
        1 & \Keyword{init}(\CounterIncAutomaton) \\[0.5ex]
        2 & \Keyword{repeat} \\
        3 & \qquad \Keyword{@every node do }
            \Keyword{for}~ i \in \range[1]{\FuncNum}
            ~\Keyword{do}~ \FuncVar_i \leftarrow \FuncVarNew_{i} \\
        4 & \qquad\Keyword{for}~ i \in \range[1]{\FuncNum}
            ~\Keyword{do}~\textit{update}(\FuncVarNew_{i}) \\
        5 & \qquad \Keyword{if}~
            \Keyword{@every node }(\forall i \in \range[1]{\FuncNum}:
            \FuncVarNew_{i} = \FuncVar_i)
            ~\Keyword{then}~ \Keyword{goto}~ 8 \\
        6 & \Keyword{until}~
            \Keyword{execute}(\CounterIncAutomaton) ~\Keyword{returns}~ \OutputNo
            \qquad
            \text{/$*$ until global counter at maximum $*$/} \\[1ex]
        7 & \Keyword{@every node do } \Keyword{for}~ i \in \range[1]{\FuncNum}
            ~\Keyword{do}~ \FuncVar_{i} \leftarrow \Self \\[0.5ex]
        8 & \Keyword{execute}(\Automaton_{\Formula[2]})
    \end{array}$
  \end{algorithm}

  \begin{algorithm}[t]
    \caption{\; $\textit{update}(\FuncVarNew_{i})$,\, as controlled by the root}
    \label{algo:update}
    $\begin{array}{r@{\quad}l}
        1' & \Keyword{for}~ \InVarReg \in \NodeSet ~\Keyword{do} \\
        2' & \qquad \OutVarRegFound \leftarrow \InVarReg;
                   ~\mathit{found}~ \leftarrow \mathit{false} \\[0.5ex]
        3' & \qquad \Keyword{for}~ \OutVarReg \in \NodeSet ~\Keyword{do} \\
        4' & \qquad\qquad \Keyword{if}~
             \Keyword{execute}(\Automaton_{\Formula_i}) ~\Keyword{returns}~ \OutputYes
             \qquad
             \text{/$*$ $\Formula_i[\InVarReg,\OutVarReg]$ satisfied $*$/} \\
        5' & \qquad\qquad\qquad\Keyword{then if}~ \mathit{found} = \mathit{false} \\
        6' & \qquad\qquad\qquad\qquad\qquad
             ~~\Keyword{then}~ \OutVarRegFound \leftarrow \OutVarReg;~
             \mathit{found} \leftarrow \mathit{true} \\
        7' & \qquad\qquad\qquad\qquad\qquad
             ~~\Keyword{else}~ \OutVarRegFound \leftarrow \InVarReg;~
             \Keyword{goto}~ 8' \\[0.5ex]
        8' & \qquad \Keyword{@}\InVarReg
              \Keyword{ do } \FuncVarNew_{i} \leftarrow \Root.\OutVarRegFound
    \end{array}$
  \end{algorithm}

  In the following,
  in the interest of clarity,
  we abstract away from many implementation details
  and only describe our construction informally.
  Recall that~$\Automaton_{\Formula}$ is
  a \emph{\kl{centralized}} \kl{automaton},
  which means that it is controlled by the root.
  The root's program is given by Algorithm~\ref{algo:pfp},
  presented as pseudo code.
  First,
  the counter is initialized to zero in Line~1
  (see below for an explanation).
  Then,
  in every iteration of the loop starting at Line~2,
  all \kl{registers}~$\FuncVar_i$ and~$\FuncVarNew_i$
  are updated in such a way that
  they represent the current and next \kl{stage},
  respectively,
  of the \kl{partial fixpoint} induction.
  For the former,
  it suffices that every \kl{node} copies,
  for all~$i$,
  the contents of~$\FuncVarNew_{i}$ to~$\FuncVar_i$ (Line~3).
  To update~$\FuncVarNew_i$,
  Line~4 calls the subroutine provided by Algorithm~\ref{algo:update}
  (which will be explained in the next paragraph).
  As a result of this algorithm,
  we have
  $\FuncVarNew_{i} =
  \FuncOperator_{\Formula[1]_i}((\FuncVar_i)_{i \in \range*{\FuncNum}})$
  for all~$i$,
  where
  $\FuncOperator_{\Formula[1]_i} \colon
  (\NodeSet^\NodeSet)^\FuncNum \to \NodeSet^\NodeSet$
  is the operator defined in Section~\ref{sec:logic}.
  Line~5 checks whether we have reached a fixpoint:
  The root asks every \kl{node} to compare, for all $i$,
  its \kl{registers}~$\FuncVarNew_{i}$ and~$\FuncVar_i$.
  The corresponding truth value is propagated back to the root,
  where $\mathit{false}$ is given preference over $\mathit{true}$.
  If the result is $\mathit{true}$,
  we exit the loop and proceed with calling~$\Automaton_{\Formula[2]}$
  to evaluate~$\Formula[2]$ (Line~8).
  Otherwise,
  we try to increment the global counter
  by executing~$\CounterIncAutomaton$ (Line~6).
  If the latter returns~$\OutputYes$,
  i.e., incrementation was possible,
  then another iteration takes place.
  However,
  if~$\CounterIncAutomaton$ returns~$\OutputNo$,
  then the counter has reached its maximum.
  This implies that we have not reached (and will not reach) a fixpoint
  so that,
  according to the \kl{partial fixpoint} semantics,
  each \kl{node} writes its own \kl{identifier}
  to every \kl{register}~$\FuncVar_i$ (Line~7)
  before the \kl{automaton} evaluates~$\Formula[2]$ (Line~8).

  Let us now describe the subprocedure $\textit{update}(\FuncVarNew_{i})$
  given by Algorithm~\ref{algo:update}.
  Using~$\RegIncAutomaton{\InVarReg}$ and~$\RegIncAutomaton{\OutVarReg}$
  provided by Proposition~\ref{prop:inc},
  the root will gradually increment its \kl{register}~$\InVarReg$ and,
  in a nested fashion,
  $\OutVarReg$ so as to let~$\InVarReg$ and~$\OutVarReg$
  range over~$\NodeSet$ (lines~$1'$ and~$3'$).
  After each increment,
  it will launch~$\Automaton_{\Formula_i}$
  to evaluate~$\Formula_i$
  with the current \kl{interpretations}
  of~$\InVarReg$ and~$\OutVarReg$ (line $4'$).
  If the result is~$\OutputYes$,
  then the root transfers the contents
  of~$\OutVarReg$ to~$\OutVarRegFound$ (Line~$6'$).
  Moreover,
  it sets the flag $\mathit{found}$ to $\mathit{true}$,
  which allows it to check
  whether the \kl{node} henceforth stored in~$\OutVarRegFound$
  is the only one to make~$\Formula_i$ true
  for the given~$\InVarReg$.
  If it is not
  (i.e., the test in Line~$5'$ eventually fails),
  we set $\OutVarRegFound = \InVarReg$.
  Finally,
  the \kl{node} whose \kl{identifier} corresponds to~$\InVarReg$
  sets its \kl{register}~$\FuncVarNew_{i}$
  to the computed value (Line~$8'$).
  To implement Line~$8'$,
  the root will send~$\OutVarRegFound$,
  along the \kl{spanning tree},
  to \kl{node}~$\InVarReg$,
  which stores it in its \kl{register}~$\FuncVarNew_{i}$.

  \paragraph*{A distributed counter.}

  We now sketch how to implement~$\CounterIncAutomaton$.
  On an \kl{underlying} \kl{graph}
  $\Graph = \tuple{\NodeSet, \EdgeSet}$
  of size $\card{\NodeSet} = n$,
  this \kl{automaton} can be used to count in a distributed manner
  from~$0$ to $n^{\CounterConstant n} - 1$.
  The basic idea is somewhat similar to the construction
  presented in Section~\ref{ssec:function-quantifiers},
  as we will identify a \kl{register valuation function}
  with a number written in base~$n$.
  More precisely,
  each of the~$n$ \kl{nodes} will have~$\CounterConstant$ \kl{registers}
  $\Register_0, \dots, \Register_{\CounterConstant-1}$,
  each storing a number between~$0$ and~$n-1$.
  Thus,
  a \kl{register valuation function}
  $\RegisterValFunc \colon
  \NodeSet \to \NodeSet^{\set{\Register_0, \dots, \Register_{\CounterConstant-1}}}$
  can be seen as a $\CounterConstant n$-digit number written in base~$n$,
  where the~$i$-th least significant digit is stored in
  \kl{register}~$\Register_{(i \bmod \CounterConstant)}$
  of the~$\floor{i / \CounterConstant}$-th \kl{node},
  with respect to some total order~$\PostOrderRel$ on~$\NodeSet$.
  In the following,
  given a \kl{spanning tree} of~$\Graph$,
  we will choose~$\PostOrderRel$ to be
  the order in which the \kl{nodes} are visited
  in the post-order traversal of the tree
  (where the children of each \kl{node}
  are visited in ascending \kl{identifier} order).
  This way,
  the~$\CounterConstant$ most significant digits are stored in the root.

  Formally,
  $\CounterIncAutomaton$ computes a \kl{transduction}
  \begin{equation*}
    \TransductionOf{\CounterIncAutomaton} \colon
    \ConfigSet{\Singleton,
               \TreeRegisterSet \cup
               \set{\Maximum, \Register_0, \dots, \Register_{\CounterConstant-1}}}
    \to
    \ConfigSet{\set{\OutputYes,\OutputNo},
               \set{\Register_0, \dots, \Register_{\CounterConstant-1}}}.
  \end{equation*}
  It expects as \kl[input configuration]{input}
  is a \kl{spanning-tree configuration}
  $\Config = \tuple{\Graph, \StateFunc, \RegisterValFunc}$
  that encodes some number $m \in \range[0]{n^{\CounterConstant n} - 1}$
  in the \kl{registers}
  $\Register_0, \dots, \Register_{\CounterConstant-1}$
  (as described above).
  In addition,
  the largest \kl{identifier} of~$\Graph$ (i.e., $n-1$)
  must be stored in each \kl{node}'s \kl{register}~$\Maximum$.
  If $m < n^{\CounterConstant n} - 1$,
  then the \kl{output configuration}~$\Config'$
  produced by~$\CounterIncAutomaton$
  is such that the new values of
  $\Register_0, \dots, \Register_{\CounterConstant-1}$
  represent the number $m + 1$
  and the root outputs~$\OutputYes$,
  indicating that the incrementation was successful.
  Otherwise,
  all \kl{registers} are set to zero
  and the root outputs~$\OutputNo$.

  In order to generate
  the first valid \kl{input configuration} for~$\CounterIncAutomaton$,
  the command $\Keyword{init}(\CounterIncAutomaton)$
  in Algorithm~\ref{algo:pfp} (Line~1)
  sets all \kl{registers}~$\Register_i$ to~$0$
  (thereby initializing the counter to zero)
  and all \kl{registers}~$\Maximum$ to $n - 1$.
  While~$0$ is already known by each \kl{node}
  (since this is the root's \kl{identifier}),
  $n - 1$ can be determined by a simple subroutine
  that is very similar to
  the \kl{automaton} described in Proposition~\ref{prop:inc}:
  every \kl{node}~$\Node[2]$ sends to its parent in the \kl{spanning tree}
  the greatest value
  among the \kl{node} \kl{identifiers} in the subtree rooted at~$\Node[2]$,
  which ensures that the largest value received by the root is $n - 1$.

  After that,
  every execution of~$\CounterIncAutomaton$
  performs incrementation by one as follows.
  First,
  the root sends the command “increment”
  in the direction of the leaf~$\Node[1]$
  that stores the least significant digits.
  More precisely,
  it sends the command to its smallest child
  (i.e., the one with the smallest \kl{identifier}),
  and every \kl{node} that receives the command
  forwards it to its own smallest child.

  Upon receiving the command “increment”,
  leaf~$\Node[1]$ checks whether at least one of its \kl{registers}
  $\Register_0, \dots, \Register_{\CounterConstant-1}$
  contains a value smaller than~$\Maximum$.
  If not,
  it sets all of them to zero
  by copying the value~$0$ from \kl{register}~$\Root$.
  Then,
  $\Node[1]$ sends the command “increment” back to its parent,
  which will forward the “carry digit” to
  the next \kl{node} in the order~$\PostOrderRel$.
  On the other hand,
  if there exists a smallest index~$i$ such that
  $\Register_i$ contains a value smaller than~$\Maximum$,
  then~$\Node[1]$ sets to zero only the \kl{registers}
  $\Register_0, \dots, \Register_{i-1}$
  and increments~$\Register_i$
  by executing a subroutine that we will describe below.
  As soon as that subroutine has terminated,
  $\Node[1]$ sends an acknowledgment to the root,
  which then instructs all other \kl{nodes} to terminate
  before switching itself to the \kl{halting state}~$\OutputYes$.

  More generally,
  whenever a \kl{node}~$\Node[2]$ receives the command “increment”,
  $\Node[2]$ tries to forward it to
  the smallest child that has not yet received it.
  (In particular,
  a command received from one child
  will be forwarded to the next smallest child.)
  If this is not possible,
  either because~$\Node[2]$ is a leaf
  or because all of its children have already sent back the command,
  then~$\Node[2]$ performs an incrementation itself.
  To do so,
  it proceeds in exactly the same way
  as described above for \kl{node}~$\Node[1]$.

  Since the root stores the most significant digit
  in its \kl{register}~$\Register_{\CounterConstant-1}$,
  it is able to detect the integer overflow
  that occurs if the input value~$m$
  is equal to $n^{\CounterConstant n} - 1$.
  In that case,
  instead of sending “increment” to its parent,
  it first instructs all other \kl{nodes} to terminate
  and then switches to the \kl{halting state}~$\OutputNo$.

  It only remains to explain the subroutine
  that allows a \kl{node}~$\Node[2]$
  to increment its \kl{register}~$\Register_i$ by one.
  To do so,
  $\Node[2]$ will send a request to the root
  including the value of \kl{register}~$\Register_i$.
  The root stores the latter in some \kl{auxiliary register}~$\Register[2]$,
  launches~$\RegIncAutomaton{\Register[2]}$ from Proposition~\ref{prop:inc},
  and then sends the result back to~$\Node[2]$,
  which writes it into~$\Register_i$.

  \medskip

  This completes the induction
  and thereby the proof of Proposition~\ref{prop:logic2aut}.
  \qed
\end{proof}

We now have the main building block
to prove the result of this section:

\begin{proof}[Proof of Proposition~\ref{prp:logic-to-automata}]
  Let~$\Formula$ be a \kl{formula} of \kl{functional fixpoint logic}
  without \kl{free variables}.
  The \kl{automaton} \kl{deciding}
  the \kl{graph property} \kl{defined} by~$\Formula$
  proceeds as follows:
  It first constructs a \kl{spanning tree}
  on the given \kl[input configuration]{input} \kl{graph}
  by executing a variant of
  the \kl{automaton} from Example~\ref{ex:spanning-tree}
  (where the root must be the last \kl{node} to enter a \kl{halting state}).
  Then,
  it launches~$\Automaton_{\Formula}$
  from Proposition~\ref{prop:logic2aut},
  with set of \kl{input registers}~$\TreeRegisterSet$.
  Finally,
  the root informs the other \kl{nodes},
  along the \kl{spanning tree},
  about the outcome,
  i.e., $\OutputYes$ or~$\OutputNo$.
  \qed
\end{proof}


\end{document}